\documentclass[aps,superscriptaddress]{revtex4-1} 
\usepackage[french,spanish,english]{babel}
\usepackage{amsmath}
\usepackage{amssymb}
\usepackage{amsthm}
\usepackage[utf8]{inputenc}
\usepackage[pdftex]{graphicx}
\usepackage{tabularx}
\usepackage{hyperref}
\usepackage[sort&compress]{natbib}
\usepackage{epstopdf}

\newcommand{\Sc}{\mathcal{S}}
\newcommand{\FmI}{\Phi_{m,I}^{(z)}}

\newcommand{\FmOJ}{\Phi_{m,O}^{(z+1)}}
\newcommand{\FmOk}{\Phi_{m,O}^{(z-1)}}
\newcommand{\FmO}{\Phi_{m,O}^{(z)}}
\newcommand{\Sb}{S_b^{(z)}}
\newcommand{\Sa}{S_a^{(z)}}
\newcommand{\Lb}{L_b^{(z)}}
\newcommand{\C}{C^{(z)}}

\newcommand{\FaO}{\Phi_{a}^{(z)}}
\newcommand{\FaOd}{\Phi_{a}^{(z+1)}}
\newcommand{\Pa}{P_{air}^{(z)}}
\newcommand{\Pt}{P_{tissue}}

\begin{document}

\title{Towards the modeling of mucus draining from human lung: role of airways deformation on air-mucus interaction.}

\author{Benjamin Mauroy}
\affiliation{Laboratoire J.A. Dieudonn\'e - UMR CNRS 7351, Universit\'e de Nice-Sophia Antipolis, Nice, France.}
\email[]{benjamin.mauroy@unice.fr}

\author{Patrice Flaud}
\affiliation{Laboratoire MSC - UMR CNRS 7057, Universit\'e Paris 7 / CNRS, Paris, France.}

\author{Dominique Pelca}
\affiliation{Physiotherapist, M2 SDE, Pierrefite, France.}

\author{Christian Fausser}
\affiliation{Physiotherapist, CHU Bic\^etre, Le Kremlin-Bicêtre, France.}

\author{Jacques Merckx}
\affiliation{Laboratoire MSC - UMR CNRS 7057, Universit\'e Paris 7 / CNRS, Paris, France.}

\author{Barrett R. Mitchell}
\affiliation{RespInnovation SAS, Seillans / Sophia Antipolis, France.}

\date{\today}

\begin{abstract}
Chest physiotherapy is an empirical technique used to help secretions to get out of the lung whenever stagnation occurs. Although commonly used, little is known about the inner mechanisms of chest physiotherapy and controversies about its use are coming out regularly. Thus, a scientific validation of chest physiotherapy is needed to evaluate its effects on secretions.

We setup a quasi-static numerical model of chest physiotherapy based on thorax and lung physiology and on their respective biophysics. We modeled the lung with an idealized deformable symmetric bifurcating tree. Bronchi and their inner fluids mechanics are assumed axisymmetric. Static data from the literature is used to build a model for the lung's mechanics. Secretions motion is the consequence of the shear constraints apply by the air flow. The input of the model is the pressure on the chest wall at each time, and the output is the bronchi geometry and air and secretions properties. 

In the limit of our model, we mimicked manual and mechanical chest physiotherapy techniques. We show that for secretions to move, air flow has to be high enough to overcome secretion resistance to motion. Moreover, the higher the pressure or the quicker it is applied, the higher is the air flow and thus the mobilization of secretions. However, pressures too high are efficient up to a point where airways compressions prevents air flow to increases any further. Generally, the first effects of manipulations is a decrease of the airway tree hydrodynamic resistance, thus improving ventilation even if secretions do not get out of the lungs. Also, some secretions might be pushed deeper into the lungs; this effect is stronger for high pressures and for mechanical chest physiotherapy. Finally, we propose and tested two adimensional numbers that depend on lung properties and that allow to measure the efficiency and comfort of a manipulation.
\end{abstract}

\maketitle

\section{Introduction}

Lung secretions are used to protect the lungs against pollutants or infections, by capturing airborne particles or cells that may enter the lungs. Secretions form an heterogenous gel-like fluid consisting mostly in water and biopolymers \cite{lai_micro-_2009}. Secretions are motioned upwards thanks to two natural mechanisms. First, the cilia located on the bronchi walls, whose motion pushes the secretions upward the bronchial tree, this phenomenon is called mucociliary clearance \cite{grotberg_respiratory_2001, king_physiology_2006, van_der_schans_bronchial_2007}). The second mechanism is cough, it involves high air flows into the bronchi \cite{basser_mechanism_1989, vandevenne_reeducation_1999, zahm_role_1991}. These mechanisms can however fail. For example, cough or cilia motion are not mature in young children, they can also been altered by pathologies or age. As a consequence, when secretions are too thick or overproduced, they can become stalled and bronchial infections may occur. Chest physiotherapy is used to treat patients whose secretions natural clearing mechanisms are altered. The techniques depend on the pathology and on the patient characteristics, typically her/his age. Bronchiolitis, asthma, chronic obstructive pulmonary disease or cystic fibrosis are typical diseases treated with dedicated chest physiotherapy techniques.  

Chest physiotherapy is based on external maneuvers on the chest. It relies on a basic concept: the pressures applied by the practitioner on the chest transmit into the lungs and deforms the bronchi. Bronchi's volume changes create an air flow inside the lung and the mechanical constraints applied on the secretions by the air flow affect the secretions and have the potential to induce motion. Secretions are gel-like materials which are motioned only if they are submitted to a stress sufficiently strong. In manual chest physiotherapy, flow and lung volume are modulated by the practitioners during the manipulations in response to the feedback they feel from the patient. Mechanical devices based on this concept have also been developed \cite{arens_comparison_1994, langenderfer_alternatives_1998, hristara-papadopoulou_current_2008}. This concept represents however a very simplified view of the lung and thorax biomechanics and chest physiotherapy remains as of today a discipline which is mostly empirical. Maneuvers are efficient but the inner mechanisms involved in their efficiency remain either unknown or non proved in a scientific way. Many chest physiotherapy techniques exist, and their definition and whether they should be used or not are decided during country wide consensus meetings. Consequently, the available pool of maneuvers can differ from one country to the other, mostly because their efficiency are difficult to compare \cite{main_conventional_2005, pryor_physiotherapy_1999}. 

Historically, most of the maneuvers were based on applying relatively strong pressures on the chest, bringing discomfort and, if ill performed, to risks to the patient, most particularly to baby and children. Thus, the use of manual chest physiotherapy for baby suffering bronchiolitis - a well known and wide spread disease - has been questioned because of the possible risks induced by the manipulations \cite{postiaux_ladite_1992, beauvois_kinesitherapie_2007}. Recently, new techniques focusing on patient comfort and using low pressures on the chest, low air flows in the lungs and induced cough have emerged and have spread quickly in physiotherapists' community. The efficiency of these new low flow techniques is however as of today still controversial, because chest physiotherapy is based on the hypothesis that secretions can be mobilized (i.e. put to motion) only with high air flows.

Chest physiotherapy has reached a state where it needs to be supported by a strong scientific background. The different questions raised by physiotherapists would benefit from a better understanding of the coupled biomechanics of chest and lungs. An integrated model of chest physiotherapy that is able to uncover the effects of hand pressures on secretions mobilization would help the practitioners to chose and define knowledgeably the techniques they are using. Most particularly, the role of the bronchial tree geometry play an important role on the air fluid dynamics \cite{mauroy_optimal_2004, mauroy_optimal_2008, mauroy_influence_2010, bokov_lumen_2010, bokov_homothety_2014} and may have an important role on secretions mobilization. Hence, in a previous paper \cite{mauroy_toward_2011}, we studied the role of the interaction between air flow, secretion thickness and bronchial tree geometry assuming it was rigid. This first theoretical frame contained some of the core elements involved in chest physiotherapy and we showed that air flow needed to be high enough for secretions to be motioned and that applying the same given air flow in the lung leads to a stationary distribution of mucus, i.e. a constant flow manipulation might see its efficiency decrease over time. These first results indicate that air flow needs to be modulated towards higher air flows in order to be able to mobilize mucus all along the manipulation. Another method for increasing the air flow induced stress might be to decrease the volume of the bronchi while keeping the air flow rate constant. This is the idea used, for example, in low volume manipulations \cite{pryor_physiotherapy_1999} or in the mechanical chest physiotherapy technique of chest compression \cite{arens_comparison_1994, langenderfer_alternatives_1998}. In order for our model to be able to capture such behaviors and to give some qualitative and comparative predictions, it was necessary to add to our previous model how lung, and its subunits bronchi and acini, deform during the manipulations. Moreover, integrating lung deformations would improve globally the quality of the predictions of our model.

Thus, in this work, we propose a new model that integrates lung deformations. The model input is an homogeneous pressure applied on the chest, and the model output is the air flow, the secretions motion and distribution and the lung configuration (airways sizes, tissue pressure, etc.). In section \ref{methods}, we describe the hypotheses of the model and in section \ref{results}, we present and discuss the predictions of the model in the case of idealized manipulations of manual chest physiotherapy and of different devices of mechanical chest physiotherapy. 

\section{Methods}
\label{methods}

This study is based on three separate sub-models: one for the lung, one for the thorax and one to mimic chest physiotherapy manipulations. We build a {\it minimal model}, which includes only the core properties we identified as playing an important role in chest physiotherapy. This choice was mainly made in order to be able to interpret correctly the interactions between these core phenomena, it was also a way to limit the computation times of the numerical studies. As a consequence, our model can perform qualitative and comparative predictions, but not quantitative predictions.

This work assumes that lung tissue mechanics and lung generations are homogeneous. These choices were made to keep the model tractable in term of physical behavior. Thus we chose to neglect phenomena such as hydrostatic pressures, inhomogeneous pressure distribution in the lung tissue or differences between airways or alveoli belonging to the same generations. The details of the model are given in the following subsections.

\subsection{Terminology.} 
\label{term}

Below is a list of definitions for some terms or variables used in this work.

\begin{description}
\item[\it{FRC}] functional residual volume, the volume of the lung at the end of a normal expiration.
\item[\it{Generation}] the generation of an airway is an index accounting for the number of bifurcations in the path from the tree root (trachea) to that airway (tree root is zero, deepest alveoli are twenty-three). We denote $z(i)$ the generation index of airway $i$.
\item[\it{Lung volume}] lung volume is denoted $V_L$ it is the sum of the tracheobronchial tree volume $V_{tbt}$ and of the acini volume $V_{ac}$.
\item[\it{Tree structure}] for an airway indexed $i$, its parent branch is indexed $p$, and its daughter branches are indexed $d_1$ and $d_2$. Although $p$, $d_1$ and $d_2$ are function of $i$, we do not make $i$ appear for the sake of notations simplification.
\item[$1 \ cmH_2O$] $1 \ cmH_2O$ is equivalent to $98.07 \ Pa$.
\item[$\FaO$] air flow in an airway in generation $z$.
\item[$\FmI$] mucus flow in an airway in generation $z$.
\item[$\Phi_{alv}^{(z)}$] air flow in an alveolus connected to an airway in generation $z$.
\item[$l_z$] length of the bronchi from generation index $z$.
\item[$\max(a,b)$] maximum function, $\max(a,b) = a$ is $a>b$ and $\max(a,b) = b$ otherwise.
\item[$n_{alv}^{(z)}$] number of alveoli connected to an airway in generation $z$ ($0$ for airways which are not alveolar ducts, $58$ for alveolar ducts, i.e. whose generation is deeper than $17$).
\item[$P_{ext}$] external chest wall pressure. It is assumed constant everywhere on the chest.
\item[$\Pa$] air pressure at mid-airway length for airway $i$.
\item[$P_{tissue}(V_L)$] mean lung tissue pressure when the volume of the lung is $V_L$. This data comes from static lung volume-pressure relationships measured by Agostoni et al. \cite{agostoni_static_2011}.
\item[$r_a^{(z)}$] air lumen area radius.
\item[$r_b^{(z)}$] airway lumen area.
\item[$\Sa$] air lumen area of bronchus $i$. Air lumen area is assumed disk shaped.
\item[$\Sb$] airway $i$ lumen area. Airway lumen area is assumed disk shaped.
\item[$V_{L,rs}(\Delta P)$] Lung volume when it is submitted to a pressure difference $\Delta P$ between the thorax and the bronchi (same pressure into the whole lung, no air movement). This quantity is built from static lung volume-pressure relationships measured by Agostoni et al. \cite{agostoni_static_2011}.
\item[$z$] generation index in the tree.
\end{description}

Note that the subscript $(z)$ corresponds to the airway generation associated to the data. In the case of properties that are applicable to any airway, we may drop this subscript for the sake of formulas simplification.

\subsection{Lung's geometry}
\label{geom}

The geometrical model of the lung used in this work is scaled to have a total lung capacity (TLC) of $6.5 \ L$, a vital capacity (VC) of $5 \ L$, a residual volume (RV) of $1.5 \ L$ and a tidal volume at rest of $0.5 \ L$ \cite{ weibel_pathway_1984,agostoni_static_2011}. According to \cite{agostoni_static_2011}, functional residual volume (FRC) is reached at $35 \ \%$ of VC, thus $FRC = RV + 0.35 \times VC$ is $3.25 \ L$.

\subsubsection{Bronchial tree model.}

The bronchial tree is modeled as an idealized airway tree with twenty-three generations. The generation index starts from generation zero (trachea) to generation twenty-two (deepest alveolar duct) and is incremented by one at each bifurcation. Although bronchial bifurcations are slightly asymmetrical \cite{tawhai_ct-based_2004, mauroy_influence_2010, florens_optimal_2011, clement_branching_2012}, we approximate the bronchial tree with a symmetric airway tree for our pioneering model to remain tractable. We assume that each airway bifurcates into two smaller identical airways, implying that all the airways belonging to a same generation are identical.

The sizes and lengths of the conducting airways (seventeen first generations) are chosen from Lambert's data \cite{lambert_computational_1982}, the different parameters are recalled in table \ref{table1}. The airways are modeled as cylindrical tubes. We assume that mucus distribution in an airway is a constant thickness layer along the cylinder wall and that mucus cannot reach the alveoli.

\begin{table}[h!]
\begin{center}
\begin{tabular}{|c|c|c|c|c|c|c|c|c|c|c|c|c|c|c|c|c|c|}
\hline
z & 0 & 1 & 2 & 3 & 4 & 5 & 6 & 7 & 8 & 9 & 10 & 11 & 12 & 13 & 14 & 15 & 16\\
\hline
l (cm) &12.00&4.76&1.90&0.76&1.27&1.07&0.90&0.76&0.64&0.54&0.47&0.39&0.33&0.27&0.23&0.20&0.17\\
\hline
d (cm) &1.671&1.1815&0.8735&0.670&0.525&0.399&0.3095&0.241&0.192&0.151&0.119&0.096&0.080&0.070&0.0615&0.055&0.0495\\
\hline
\end{tabular}
\caption{Lung's data for the geometry of the conductive airways at FRC $=1.75 \ L$ (end expiration volume), from Lambert et al \cite{lambert_computational_1982}: $z$ is the generation index; $l$ is the airway length; $d$ is the airway diameter.}
\label{table1}
\end{center}
\end{table}

\subsubsection{Alveoli and acini geometrical models.}

Alveoli are assumed to be spherical and to stem from airways walls from the seventeenth generation.
The total number of alveoli $N_a$ is documented in the literature \cite{ochs_number_2004} and we used $N_a = 480 \ 10^6$. 
FRC volume can be decomposed as the volume of the bronchial tree plus the volume of the alveoli. Our model computations give a FRC volume for the bronchial tree of about $0.67 \ L$ and a FRC total alveoli volume of about $2.58 \ L$. Consequently, the volume of an alveolus at FRC used in our model is $v_a = 5.37 \ 10^6 \ \mu m^3$ (diameter of $217 \ \mu m$). This value is slightly higher than the value measured in \cite{ochs_number_2004} ($4.2 \ 10^{6} \ \mu m^3$, i.e. a diameter of $200 \ \mu m$), the discrepancy may arise from the measures being done in a lung's volume that is not FRC. The exchange surface in the lungs consists in the total alveoli surface and can be estimated in our model for FRC to $\sim 70 \ m^2$. 

Alveoli are stemming from bronchi walls in the acini units, that corresponds to the last sixth generations of the lungs. In our model, each acinus has $2^6 - 1 = 63$ alveolar ducts.
On the contrary of bronchi in the seventeen first generations whose diameters and lengths decrease when generation index increases, the bronchi in the acini, called alveolar ducts, keep roughly similar size whatever their generation \cite{weibel_pathway_1984}. Thus, we assume that the number of alveoli is homogeneously distributed on the alveolar ducts walls, and to reach a total of $480 \ 10^6$ alveoli, the amount of alveoli per alveolar ducts in the acini is $58$, a bit higher than the range from $20$ to $40$ reported in \cite{whimster_microanatomy_1970}. It remains however fully in the range of physiological patterns, since there is quite large variation between individuals, as reported for example in \cite{ochs_number_2004}.

\subsection{Lung's mechanics}

\begin{figure}[h!]
\includegraphics[height=5cm]{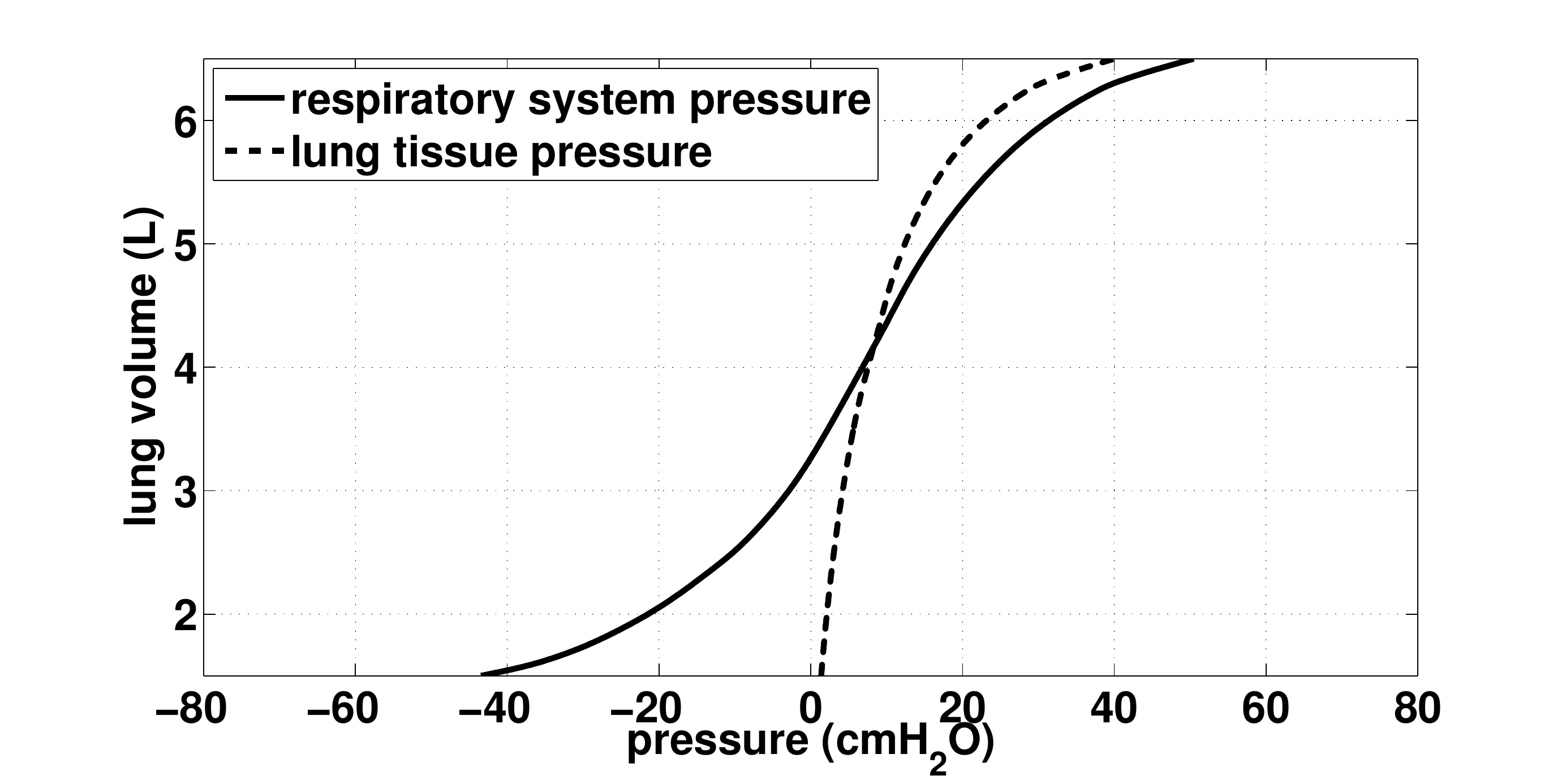}
\caption{Data (static) from Agostoni et al. \cite{agostoni_static_2011}. Continuous curve: total respiratory lung volume-pressure relationship, used in our work to model acini mechanics. Dashed curve: Lung tissue pressure, used in our work to compute bronchi transmural pressure.}
\label{agostoniP}
\end{figure}

Our goal is to be able to compute the volume of the airways and alveoli as a function of two input pressures: the pressure applied on the thorax $P_{ext}$ and the fluid pressures inside the airways $P_{air}^{(z)}$ ($z$ is the generation index). In our model, the pressure applied on the thorax may either come from muscles action, like in ventilation, or from chest physiotherapy manipulation; the pressure inside the airways is a consequence of the fluid mechanics in the tree. For the sake of simplification, we assume that the pressure in the lung tissue $\Pt$ to be homogeneous throughout the lung. We assume that lung tissue pressure $\Pt$ depends on the lung volume only and is computed from the static volume-pressure relationship from Agostoni et al work \cite{agostoni_static_2011,dangelo_statics_2005}, see figure \ref{agostoniP} (dashed curve). Similarly, we assume that bronchi and alveoli mechanics do not depend on their location in the lung.

To model the lung mechanics, we divided the tree in two parts: the conductive airways and the acini. Mechanics of airways is based on the transmural pressure-airways section relationships proposed by Lambert et al \cite{lambert_computational_1982}. Mechanics of the acini is based on the total respiratory system lung volume-pressure relationship from Agostoni et al \cite{agostoni_static_2011,dangelo_statics_2005}, see figure \ref{agostoniP} (continuous curve).

\subsubsection{Conductive airways mechanics (up to the 17th generation).} 

\begin{figure}[h!]
\includegraphics[height=5cm]{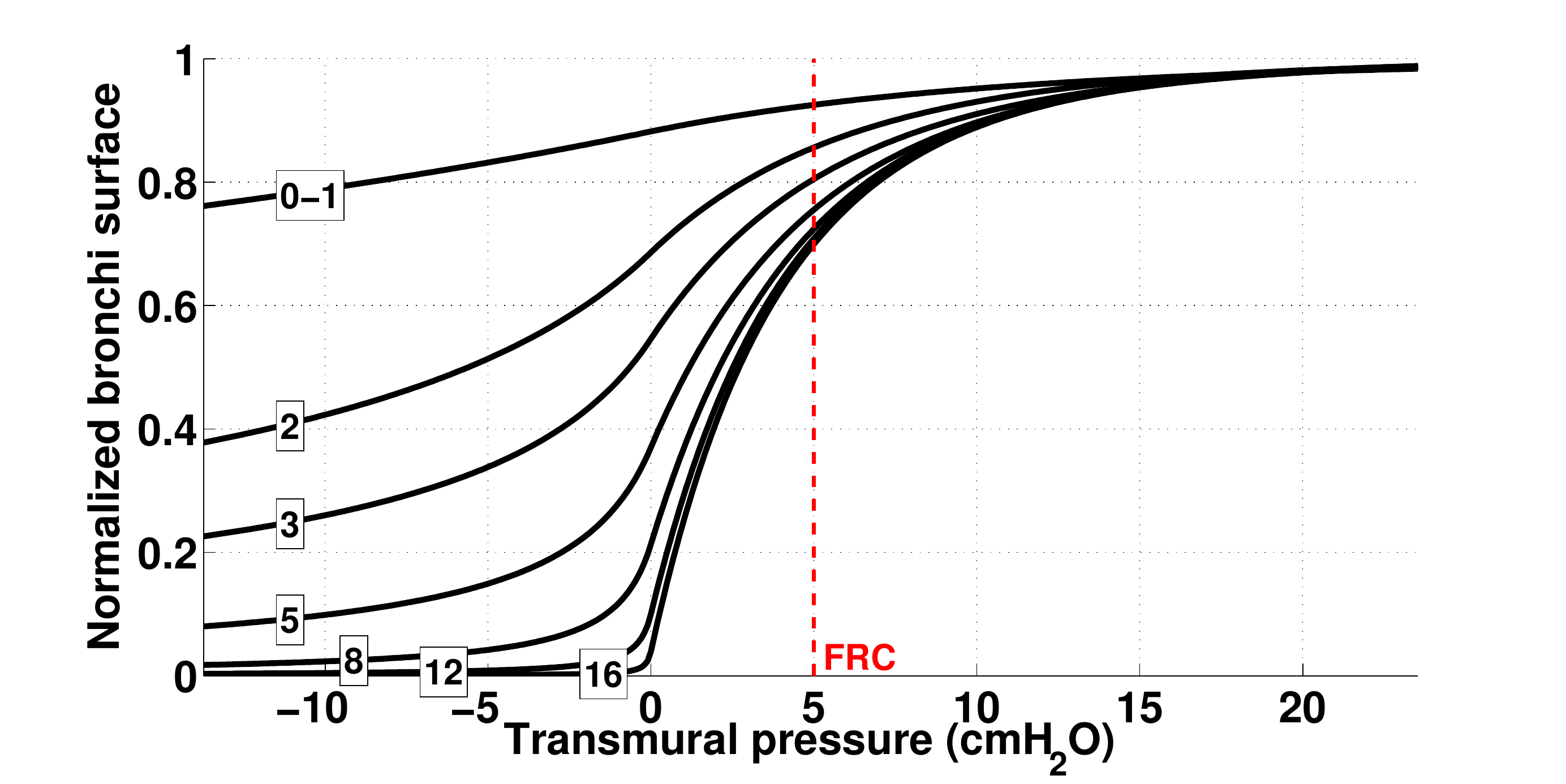}
\caption{Dependence of bronchi surface area (normalized) with bronchi transmural pressure. The numbers correspond to the generation index and the red dashed line represents bronchi state at FRC.}
\label{lambertPlot}
\end{figure}

To compute airways diameters, we used data from Lambert et al. \cite{lambert_computational_1982}. Lambert et al built data based static relationships of bronchi's airways lumen area versus their transmural pressure for the seventeenth first generations. Their estimations are given per generation. For an airway belonging to generation $z$ with a transmural pressure $\Delta P$, they proposed that the airway lumen area $S_z(\Delta p)$ is a sigmoid function of transmural pressure:			

\begin{equation}
\label{lambert1982}
\Sc_z(\Delta P) = \left\{
\begin{array}{ll}
\alpha_0(z) \left(1 - \frac{\Delta P}{P_1(z)} \right)^{-n_1(z)} A_m(z) & \text{when $\Delta P \leq 0$}\\
\left(1-(1-\alpha_0(z)) (1-\frac{\Delta P}{P_2(z)})^{n_2(z)}\right) A_m(z)  & \text{when $\Delta P > 0$}
\end{array}
\right.
\end{equation}

\noindent with $P_1(z) = \frac{\alpha_0(z) n_1(z)}{\alpha'_0(z)}$ and $P_2(z) = -n_2(z) \frac{1 - \alpha_0(z)}{\alpha'_0(z)}$. $A_m$ represents the maximal possible lumen area. The quantity $\alpha_0 A_m$ is the airway area when transmural pressure is $0$. At functional residual regime (FRC, end of normal expiration), the transmural pressure is about $500 \ Pa$. The values used for the different parameters in the previous formulas are detailed in table \ref{table2}; the dependence of the bronchi surface area with transmural pressure are plotted on figure \ref{lambertPlot}. 

\begin{table}[h!]
\begin{center}
\begin{tabular}{|c|c|c|c|c|c|c|c|c|c|c|c|c|c|c|c|c|c|}
\hline
z & 0 & 1 & 2 & 3 & 4 & 5 & 6 & 7 & 8 & 9 & 10 & 11 & 12 & 13 & 14 & 15 & 16\\
\hline
$\alpha_0$ &0.882&0.882&0.686&0.546&0.450&0.370&0.310&0.255&0.213&0.184&0.153&0.125&0.100&0.075&0.057&0.045&0.039\\
\hline
$ \alpha'_0$ ($\times 10^{-3} \ Pa^{-1}$) &1.1&1.1&5.1&8.0&10.0&12.5&14.2&15.9&17.4&18.4&19.4&20.6&21.8&22.6&23.3&23.9&24.3\\
\hline
$n_1$ &0.5&0.5&0.6&0.6&0.7&0.8&0.9&1.00&1.00&1.00&1.00&1.00&1.00&1.00&1.00&1.00&1.00\\
\hline
$n_2$ &10.00&10.00&10.00&10.00&10.00&10.00&10.00&10.00&10.00&10.00&10.00&9.00&8.00&8.00&8.00&7.00&7.00\\
\hline
$A_m$ (cm$^2$) &2.37&2.37&2.80&3.50&4.50&5.30&6.50&8.00&10.20&12.70&15.94&20.70&28.80&44.50&69.40&113.00&180.00\\
\hline
\end{tabular}
\caption{Lung's data for the static mechanical behavior of the conductive airways, from Lambert et al \cite{lambert_computational_1982}. $z$ is the generation index, see equation (\ref{lambert1982}) for the definitions of the other numbers $\alpha_0$, $\alpha'_0$, $n_1$, $n_2$ and $A_m$.}
\label{table2}
\end{center}
\end{table}

Airways transmural pressure is the difference between the pressure in the lung's tissues around the bronchi and the bronchi inner pressure due to fluid mechanics in the lung. The fluid mechanics in the airway tree is detailed below. We assume that tissue pressure depends on the volume of the lungs only; it is computed from the volume - lung pressure  static relationship measured by Agostoni et al \cite{agostoni_static_2011}. Using our terminology, the lumen area $\Sb$ of a bronchi in generation $z$ ($z \leq 16$) is

\begin{equation}
\label{lumenAreaEq}
\Sb = \Sc_{z} \left(\Pt (V_L)-P_{air}^{(z)} \right)
\end{equation}

We recall that $\Pt(V_L)$ is given by the volume-pressure relationship from Agostoni et al. \cite{agostoni_static_2011}, see figure \ref{agostoniP} (dashed curve). To simplify, we will assume that $P_{air}^{(z)}$ is evaluated in the middle of the bronchus $z$. Moreover, the pressure - lumen area relationships assumes that airway walls are always at equilibrium with their corresponding transmural pressure, thus bronchi mechanics in our model is quasi-static. From these laws, we can derive the volume of the tracheobronchial tree,
\begin{equation}
\label{tbtVolEq}
V_{tbt}(V_L, P_{air}) = \sum_{z=0}^{16} 2^z \Sc_{z} \left( \Pt(V_L)-P_{air}^{(z)} \right) \Lb
\end{equation}

\subsubsection{Alveolar ducts and alveoli mechanics (last six generations).} 

The last sixth generations of the lung correspond to the acini. Each acinus in our model consists in $2^6-1 = 63$ alveolar ducts and each alveolar duct is connected to $58$ alveoli (see model geometry section). An alveolar duct with its corresponding $58$ alveoli will be referred to as {\it an alveolar duct unit}. There is $2^{17} \times 63$ alveolar duct units in our model. Lambert et al data does not cover those generations, thus we used another approach to approximate the mechanical behavior of an alveolar duct unit.

The pressure difference between the inside of an alveolar duct unit in generation $z$ and the body surface is $P_{air}^{(z)} - P_{ext}$. Consequently, the volume the lung would have under static condition with the same pressure difference {\it in the whole tree} is given by the respiratory system pressure - volume relationship from \cite{agostoni_static_2011}, see continuous curve on figure \ref{agostoniP}. However, the lung volume measured by Agostoni et al includes the tracheobronchial tree volume, thus to reach the volume of acini only, it is necessary to subtract from the lung volume the volume of the tracheobronchial tree as it is in the configuration of Agostoni et al measurements: static and the glottis open. As a consequence, the reference pressure (= atmospheric pressure) spans the whole tree. Hence, the volume of an alveolar duct unit in generation index $z$ in the tree is:

\begin{equation}
\label{acinVolEq}
v_{adu}\left(P_{air}^{(z)}, P_{ext}\right) = \frac1{2^{17} \times 63} \left( V_{L,rs} ( P_{air}^{(z)} - P_{ext}) - V_{tbt}\left(V_{L,rs} ( P_{air}^{(z)} - P_{ext} ),0\right) \right)
\end{equation}

Now, to compute alveolar ducts and alveoli volume, we assume their respective volume ratio in an alveolar duct unit remains unchanged whatever the lung volume. We assume the length of alveolar ducts to be $L_{ad} = 0.7 \ mm$ \cite{haefeli-bleuer_morphometry_1988}. The quantity $\alpha$ represents the volume ratio of an alveolar duct into an alveolar duct unit. Hence, for airways in a generation $z$ in the acini ($z > 16$):

\begin{equation}
\label{acinMeca}
\left\{
\begin{array}{ll}
\Sb = \alpha v_{adu}\left(P_{air}^{(z)}, P_{ext}\right) / L_{ad} & \text{ (alveolar duct lumen area)}\\
v_{alv}^{(z)} = (1-\alpha) v_{adu}\left(P_{air}^{(z)}, P_{ext}\right) & \text{ (alveolus volume)}
\end{array}
\right.
\end{equation} 

\noindent We compute $\alpha$ from FRC state and assume it is a constant: $\alpha = 0.17$. Finally, we can compute the total volume of the acini in the tree,
\begin{equation}
V_{ac}(P_{air}, P_{ext}) = \sum_{z=17}^{22} 2^z v_{adu}\left( P_{air}^{(z)}, P_{ext} \right) 
\end{equation}

\subsubsection{Computation of the lung volume $V_L$.}

To compute the conductive airways lumen areas with equation (\ref{tbtVolEq}), we need to compute the volume of the lung $V_L$. Acini volume is known through equation (\ref{acinVolEq}) and lung volume is the sum of acini volume and of conductive airways volume. Consequently, lung volume $V_L$ depends on $P_{air}$ and $P_{ext}$ and is the solution of the equation:

\begin{equation}
\label{volEq}
V_L = V_{ac}(P_{air}, P_{ext}) + V_{tbt}(V_L, P_{air})
\end{equation}

Once computed, lung volume $V_L$ is used to compute the tracheobronchial tree lumen areas using equation (\ref{lumenAreaEq}). Consequently, the airway lumen area of generations $z \leq 16$ expressed in equation (\ref{lumenAreaEq}) can be reformulated as a function of $P_{ext}$ and $P_{air}$: $\Sb = \Sc_{z} \left(P_{tissue}\left(V_L(P_{ext}, P_{air})\right)-P_{air}^{(z)} \right)$. Merging with the equation on the first line of equations (\ref{acinMeca}), one can summarize the airway mechanics into a vectorial form:
\begin{equation}
\label{airwayMeca}
\mathcal{S} = H(P_{ext}, P_{air})
\end{equation}
\noindent with 
$$
H(P_{ext}, P_{air})^{(z)} =  \left\{
\begin{array}{ll}
\Sc_{z} \left(P_{tissue}\left(V_L(P_{ext}, P_{air})\right)-P_{air}^{(z)} \right) & \text{ for } 0 \leq z \leq 16\\
\alpha v_{adu}\left(P_{air}^{(z)}, P_{ext}\right) / L_{ad} & \text{ for } 17 \leq z \leq 22
\end{array}
\right.
$$

\subsection{Air and mucus mechanics.}

In the previous section, we defined a model that mimics how airways and alveoli are behaving when they are submitted to an external pressure $P_{ext}$ and an internal fluid pressure $P_{air}$.
Whereas the external pressure $P_{ext}$ is an input of the model, fluid pressures in the airways $P_{air}$ are outputs. Fluid pressures and flows come from the interactions between airways deformations and fluid mechanics of air and mucus.

We will distinguish two lumen areas: the bronchi lumen area, whose surface is measured by the quantity $\Sb$ in generation $z$, and the air lumen area, whose surface is measured by the quantity $\Sa$ in generation $z$. The difference $\Sb - \Sa$ represents the surface of the airway section filled with mucus.

\subsubsection{Air mechanics.}

In this section, we will derive the equations on the air lumen area surface $\Sa$. Air is assumed incompressible and Newtonian. Its density is $\rho_a = 1 \ kg.m^{-3}$ and its viscosity is $\mu_a = 1.8 \ 10^{-5} \ Pa.s$. We assume that air motion in a bronchus is at low regime with fully developed profiles as in \cite{mauroy_toward_2011}. The rate of air volume change in an airway is equal to the air flow getting in the airway - or getting out the airway, in which case we consider there is a negative airflow getting in the airway. Air flow getting in an airway in generation $z$ comes from its two daughter branches standing in generation $z+1$. The conservation law for air in generation $z$ is:
\begin{equation}
\label{fm1}
\frac{d \Sa}{dt} \Lb = - \left(2 \FaOd  - \FaO\right) - n_{alv}^{(z)} \Phi_{alv}^{(z)}
\end{equation}
The minus sign means that air flow coming in the tree is modeled as negative. Note that we could also have computed the balance with the air flow coming in from the parent branch instead of the daughter branch, however this would lead to the similar equations once rewritten at the tree level. The number $n_{alv}^{(z)}$ corresponds to the number of alveoli connected to an airway in generation $z$, it is equal to $0$ if the airway belongs to generations zero to sixteen and is equal to $58$ if it belongs to a deeper generation, see geometry section upwards. $\Phi_{alv}^{(z)}$ corresponds to the part of the airway air flow that is induced by one alveolus belonging to generation $z$. $\Phi_{alv}^{(z)}$ is equal to the rate at which the volume $v_{alv}^{(z)}$ changes in time. It is computed from equation (\ref{acinMeca}): 
\begin{equation}
\label{fm2}
\frac{d v_{alv}^{(z)}}{dt} = - \Phi_{alv}^{(z)} 
\end{equation}
Because of our hypotheses, the pressure drop per unit length in an airway, called $C=\frac{\partial p}{\partial z}$ depends on the generation index only. $\C$ in an airway of generation $z$ can be computed from the air flow $\FaO$ in the bronchus and from air and mucus lumen areas, respectively denoted $\Sa$ and $\Sb$:
\begin{equation}
\label{fm3}
\C = F(\FaO, \Sa, \Sb)
\end{equation}
$F$ computation is based on low regime and fully developed fluid dynamics hypotheses, as in our previous work \cite{mauroy_toward_2011}. Details about how $F$ is computed are recalled in Appendix \ref{AppF}. Since pressure reference is that of atmospheric pressure, the pressure at the root of the tree (trachea) is zero. The pressure drop in an airway of generation $z$ is $C^{(z)} \Lb$ and the pressure $\Pa$ in generation $z$ is the sum of the pressure drops of the airways connecting the tree root (trachea) to a generation $z$ airway:
\begin{equation}
\label{fm4}
\Pa =  \C \frac{\Lb}2 + \sum_{g=0}^{z-1} C^{(g)} L_b^{(g)}
\end{equation}
The $1/2$ in the first term means that $\Pa$ is air pressure at mid-bronchus length.

\subsubsection{Mucus mechanics.} 

As in \cite{mauroy_toward_2011}, we assume that secretions are a Bingham fluid: it is solid if its inner shear stress $\sigma$ is lower than a stress threshold $\sigma_0$ and liquid with viscosity $\mu_m$ otherwise. In this work, we assume $\sigma_0 = 0.1 \ Pa$ and $\mu_m = 0.1 \ Pa.s$, see \cite{mauroy_toward_2011}.  Secretions density is that of water, i.e. $1000 \ kg/m^3$.
Mucus is motioned by air flow when the shear forces induced by air flow on mucus exceed the shear threshold $\sigma_0$. Mucus can have three states: either fully solid, fully liquid or both solid/liquid. The state of mucus depends on a radius $r_0 = \left| 2 \sigma_0 / C \right|$ where $C$ is the pressure drop per unit length in the branch \cite{mauroy_toward_2011}. Mucus is solid if $r_0 > r_b$, liquid if $r_0 < r_a$. If $r_a < r_0 < r_b$, then mucus is solid between radius $r_a$ and $r_0$ and liquid between radius $r_0$ and $r_b$.

Mucus volume change in an airway in generation $z$ is the balance between mucus input $\FmI$ and output $\FmO$. 
Mucus volume in a generation $z$ airway is the difference between airway total volume and the volume occupied by air: $\Sb \Lb - \Sa \Lb$.
Since the airway length $L_b$ is assumed constant, the time derivative of this equation leads to
the conservation law for mucus: 
\begin{equation}
\label{mm1}
\left(\frac{d \Sb}{dt} - \frac{d \Sa}{dt}\right) \Lb = |\FmI| - |\FmO|
\end{equation}
The flow of mucus $\FmO$ that gets out of an airway through its interaction with air flow can be computed from air properties in the airway. Mucus flow $\FmO$ in an airway in generation $z$ is a function of the pressure drop per unit length in the airway $\C$ and of the bronchi and air lumen areas of the airway:
\begin{equation}
\label{mm2}
\FmO = G(\C, \Sa, \Sb) 
\end{equation}
Details on how to compute $G$ are given in Appendix \ref{AppG}, the method is the same as in \cite{mauroy_toward_2011}. Mucus flow in airways occurs only if the local shear stress induced by air flow on mucus overcomes the mucus yield stress $\sigma_0$. The flow of mucus that gets in a generation $z$ airway comes either from the two daughter airways in generation $z+1$ or from the parent airway in generation $z-1$, depending on mucus flow direction. This is reflected by the sign of mucus flow:
\begin{equation}
\label{mm3}
\FmI = \frac12 \max\left(0,\FmOk\right) - 2 \min\left(0,\FmOJ\right)
\end{equation}
If mucus flows $\FmOJ$ in the two daughter airways (generation $z+1$) are negative, then mucus is going up the tree and thus accounts twice for mucus inflow in an airway of generation $z$. If mucus in the parent airway (generation $z-1$) is positive, then mucus in that airway is going down and accounts half for mucus inflow in an airway of generation $z$. If the flow in one of these neighboring airways are of the opposite sign as stated, then it does not account for the mucus inflow in an airway of generation $z$. 

\subsection{Solving equations.}

The model consists in the set of equations (\ref{airwayMeca}), (\ref{fm1}), (\ref{fm2}), (\ref{fm3}), (\ref{fm4}), (\ref{mm1}), (\ref{mm2}), (\ref{mm3}). The system of equations is implemented numerically in {\it C++} using a vectorial formulation based on {\it Eigen 3.2.1} (\href{http://eigen.tuxfamily.org/}{http://eigen.tuxfamily.org/}) and {\it OpenMP} (\href{http://openmp.org/}{http://openmp.org/}) multiprocessing. It is solved thanks to a full implicit first order method for differential equations (\ref{fm1}), (\ref{fm2}) and (\ref{mm1}). A Newton method is used to solve equation (\ref{volEq}) in order to compute the function $H$, see equation (\ref{airwayMeca}). Details of the mathematical and numerical implementations are given in Appendix \ref{numerics}.

In order to be able to capture the effects of a $20 \ Hz$ oscillating external pressure, the typical time step in our simulations is $5.10^{-3} \ s$. The time needed to compute one time step on $6$ cores Xeon 5645 ranges from $0.1$ to $0.2 \ s$. In this study, the typical manipulation mimicked by our model covers about $200 \ s$, with a total computation time of $4000$ to $8000$ seconds.

\section{Results/Discussion}
\label{results}

The model we describe in the previous sections allows to test how the changes in airway diameters affect air and mucus distributions in the tree. The model contains the minimal physics that we hypothesize to be main drivers of chest physiotherapy: the tree structure, linear air fluid mechanics, Bingham fluid mechanics to model secretions behavior and quasi-static deformable airways. In this section, we present the model predictions with pressure inputs mimicking chest physiotherapy. In section \ref{inits}, we define the initial state which is used in all the next simulations. This initial state is not representative of any specific pathology, it only reflects a lung with an excess of secretions near the eighth generation. In section \ref{manuPhysio}, we present how our model is responding to inputs that mimic manual chest physiotherapy. Next, in section \ref{mechaPhysio} we present how our model is responding to inputs that mimic high frequency chest wall oscillations. We study most particularly the role of an added static pressure on the chest. In order to be able to compare both the efficiency and comfort of these manipulations we define in sections \ref{shrek} and \ref{weasel} two numbers that quantify manipulations efficiency and comfort. 

\subsection{Mimicking chest physiotherapy manipulations}
\label{inits}

We will test different input pressures mimicking manual or mechanical chest physiotherapy. We recall that our model assumes the pressure to be homogeneous on the thorax. For each manipulation, we will focus on three quantity computed from the model outputs:
\begin{itemize}
\item the quantity of mucus expelled out of the tree $v_{out}$ - the net reduction of mucus volume in the lung after a manipulation.
\item the relative change in hydrodynamic resistance $r$ during and after a manipulation - this quantity is correlated to the feeling for the patient to breath easily or not.
\item the mean mucus position in the tree, that reflects how deep mucus stands in the lung. It is equal to 
$$
mmp = \frac{- v_{out} + \sum_{z=0}^N \left( z \ 2^z (\Sb - \Sa) \Lb \right) }{v_{out} + \sum_{z=0}^N \left(2^z (\Sb - \Sa) \Lb \right)}
$$
\end{itemize}

\begin{figure}[h!]
A
\includegraphics[height=3cm]{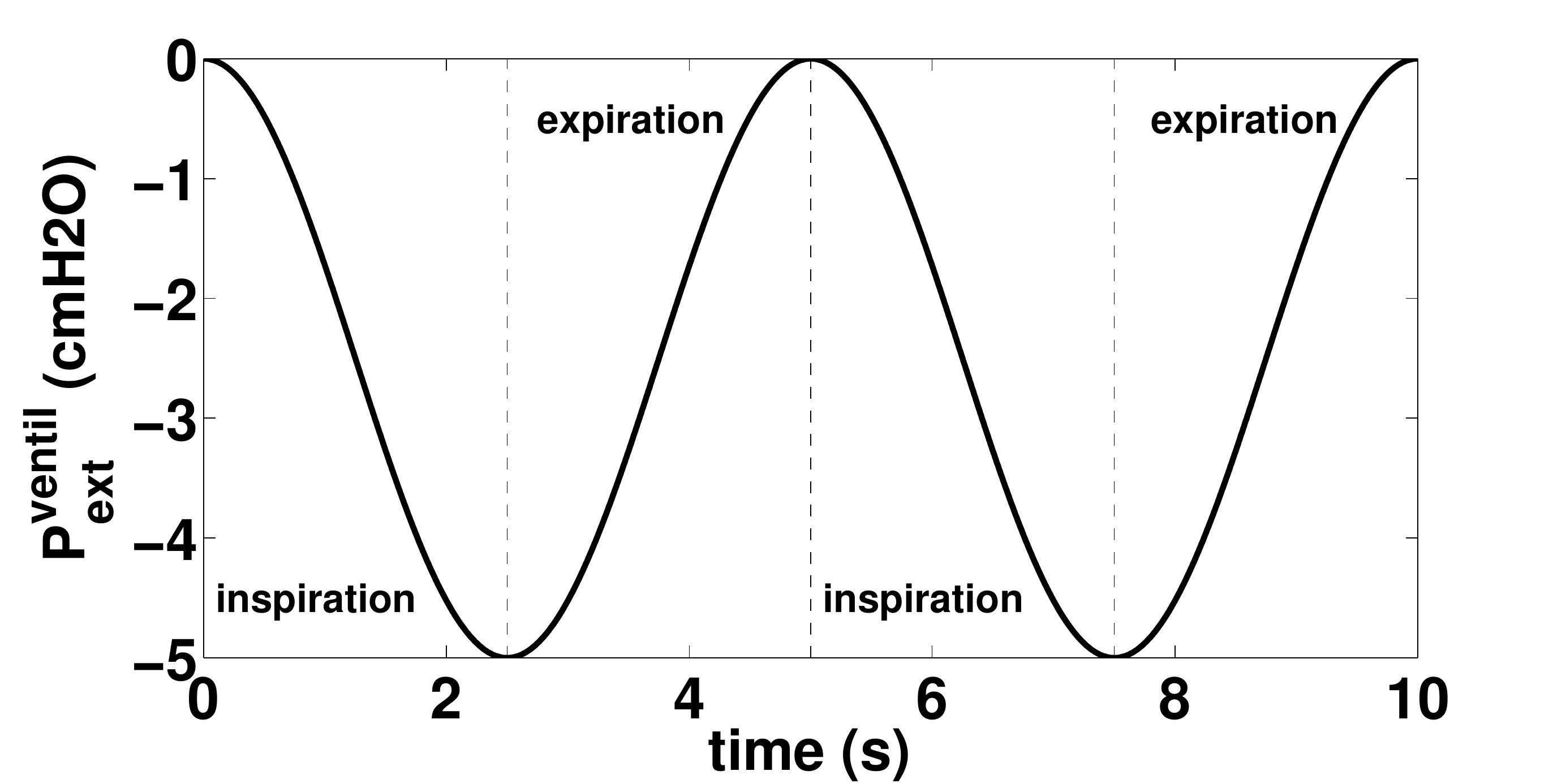}
B
\includegraphics[height=3cm]{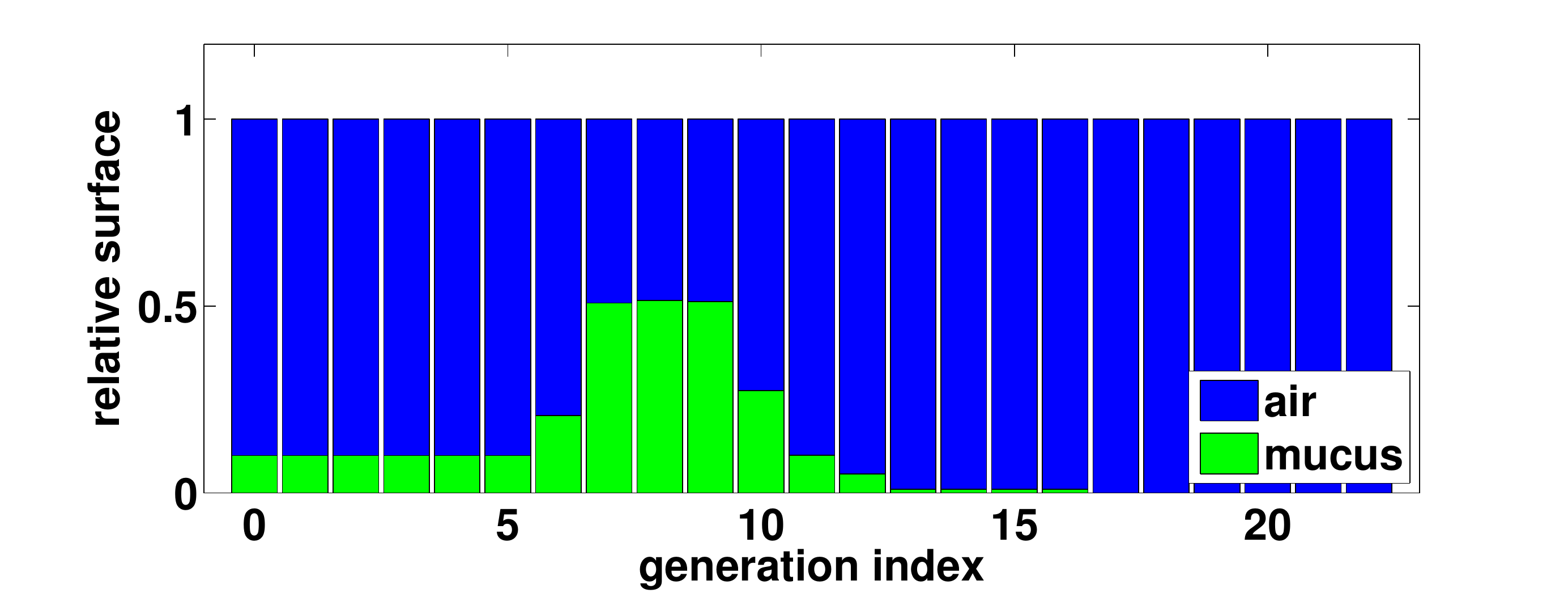}
\caption{A: External oscillating pressure to model normal rest ventilation. B: Initial distribution of mucus in the tree: air is plotted in blue and mucus in green, bars height represents a proportion of airway surface at FRC. This distribution is not affected by normal rest ventilation.}
\label{initialMucDist}
\end{figure}

All numerical experiments are performed with a background chest motion that corresponds to an idealized normal ventilation of the patient. We model normal ventilation as a sinusoidal negative pressure applied on the thorax with a frequency of $0.2 \ Hz$ and an amplitude $P_v = - 5 \ cmH_2O$, see figure \ref{initialMucDist}A: 
$$
P_{ext}^{ventil}(t) = P_v  \frac{\left(1-\cos(2 \pi t / 5)\right)}{2}$$
With such data as an input, our model predicts a tidal volume of $0.5 \ L$ ($6 \ L/min$). This value for tidal volume is typical of rest regime ventilation \cite{weibel_pathway_1984}.

The initial mucus distribution in the tree is plotted on figure \ref{initialMucDist}B. It is homogenous in the first six generations and fills about ten percents of the bronchi lumen area. Then mucus thickness increases in the three next generations to reach its maximal thickness in generation eight, filling about fifty percents of the bronchi lumen area. Finally, mucus thickness decreases and reaches zero percent in generation sixteen. The initial configuration for mucus has been chosen in such a way mucus is almost not affected by normal rest ventilation.

\subsection{Manual physiotherapy}
\label{manuPhysio}

\begin{figure}[h!]
A \includegraphics[height=3cm]{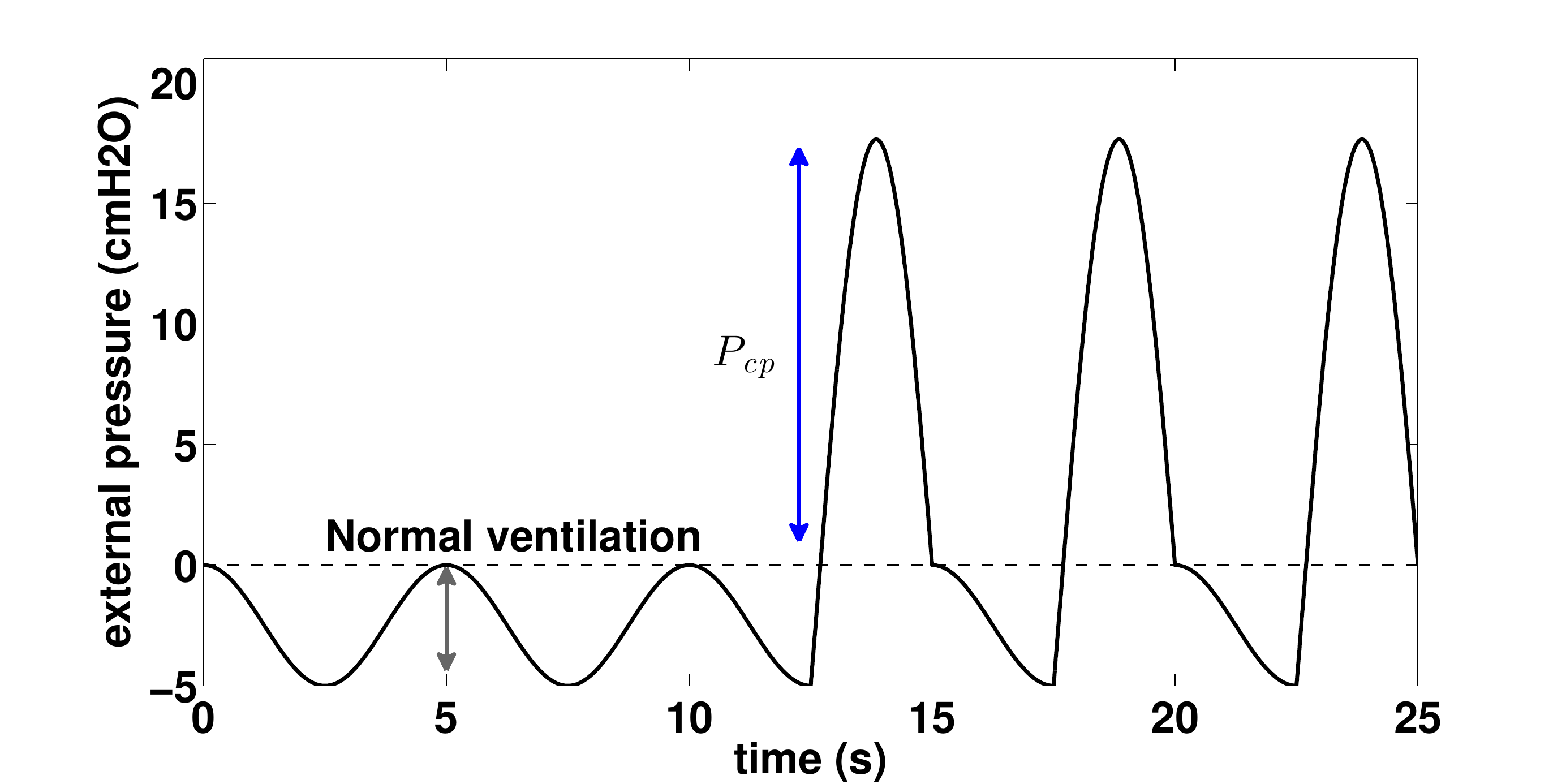}
B \includegraphics[height=3cm]{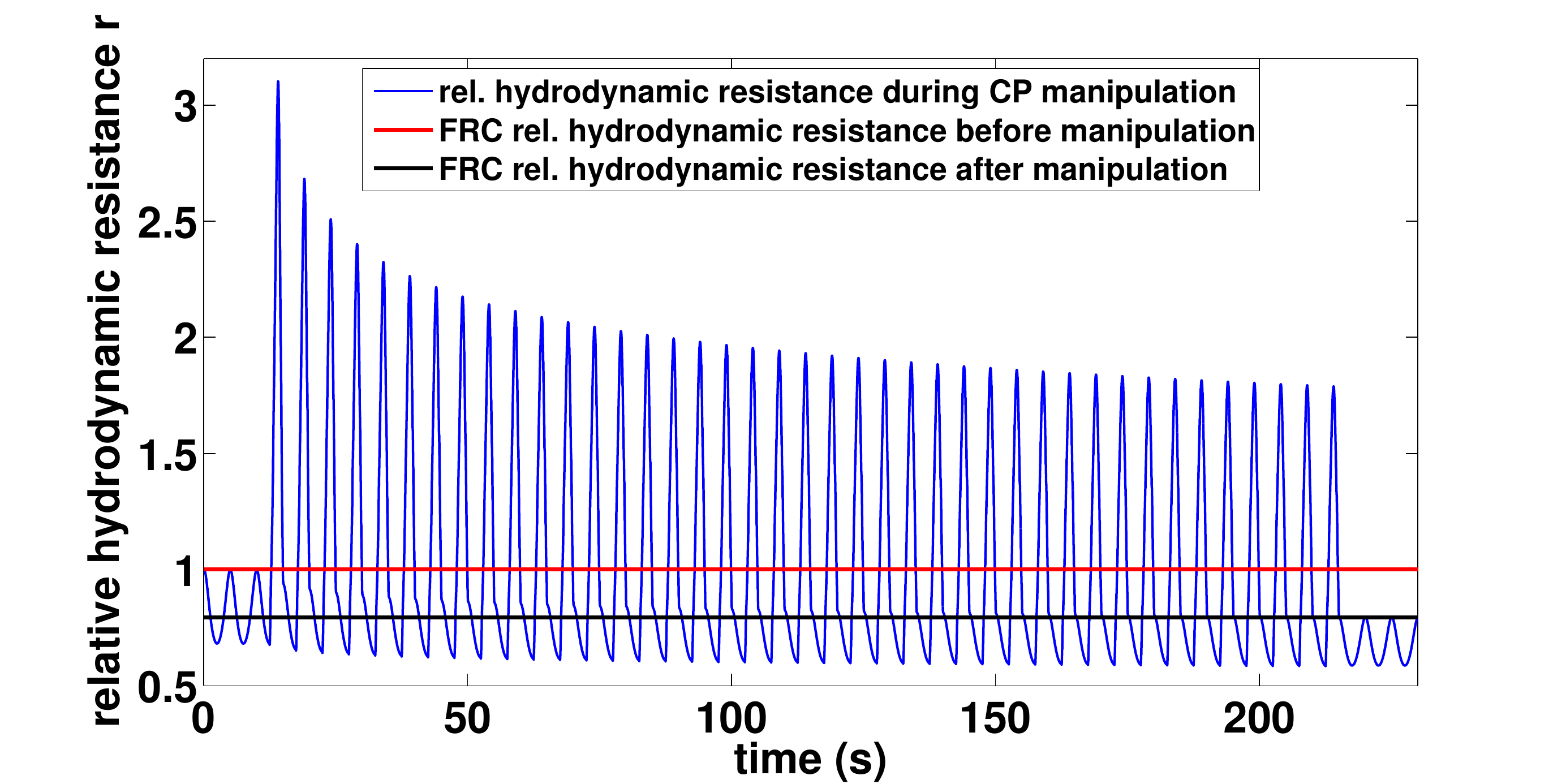} 
\caption{A: Chest pressure $t \rightarrow P_{ext}^{manual}(t)$ used as an input for our model to mimic manual chest physiotherapy. Pressure is assumed homogeneous on the thorax, case with $P_{cp} = 20 \ cmH_2O$. B: Relative hydrodynamic resistance $r$ variations during a manipulation with $P_{cp} = 20 \ cmH_2O$ (blue). The red line corresponds to FRC relative hydrodynamic resistance before the manipulation ($r_{|t=0s}=1.00$), and the black line corresponds to FRC relative hydrodynamic resistance at the end of the manipulation ($r_{|t=230s}=0.79$).}
\label{ManualPextRes}
\end{figure}

Manual chest physiotherapy is performed by applying pressures on the chest with hands. The pressures are applied during the expiration phase only and practitioners are guided by chest motion and reaction to their manipulations. Although the hand locations play an important role in chest physiotherapy, our model accounts only for pressures amplitude and time dependence: pressure is applied homogeneously onto the chest. We mimicked chest physiotherapy in our model by altering the ventilation signal during expiration, as shown on figure \ref{ManualPextRes}A:
$$
P_{ext}^{manual}(t) = P_{ext}^{ventil}(t) + P_{cp} \max \left( \sin(2 \pi t / 5) , 0 \right)
$$

Manipulation pressures are applied regularly during one session of $230$ seconds, except for the $10$ first and last seconds, where only normal ventilation occurs. The pressures applied are all the same during the whole manipulation. The twenty-five first seconds of pressure input are plotted on figure \ref{ManualPextRes}A with $P_{cp} = 20 \ cmH_2O$. In manual chest physiotherapy, the pressure is not applied on the whole chest at once. The choice of the location for pressure application is made by the physiotherapist in order to optimize the pressure effect. Our model is for now not able to account for pressure spatial inhomogeneities and as a first approximation, we assume the hand pressure spreads over the whole chest homogeneously.

\begin{figure}[h!]
A \includegraphics[height=2.5cm]{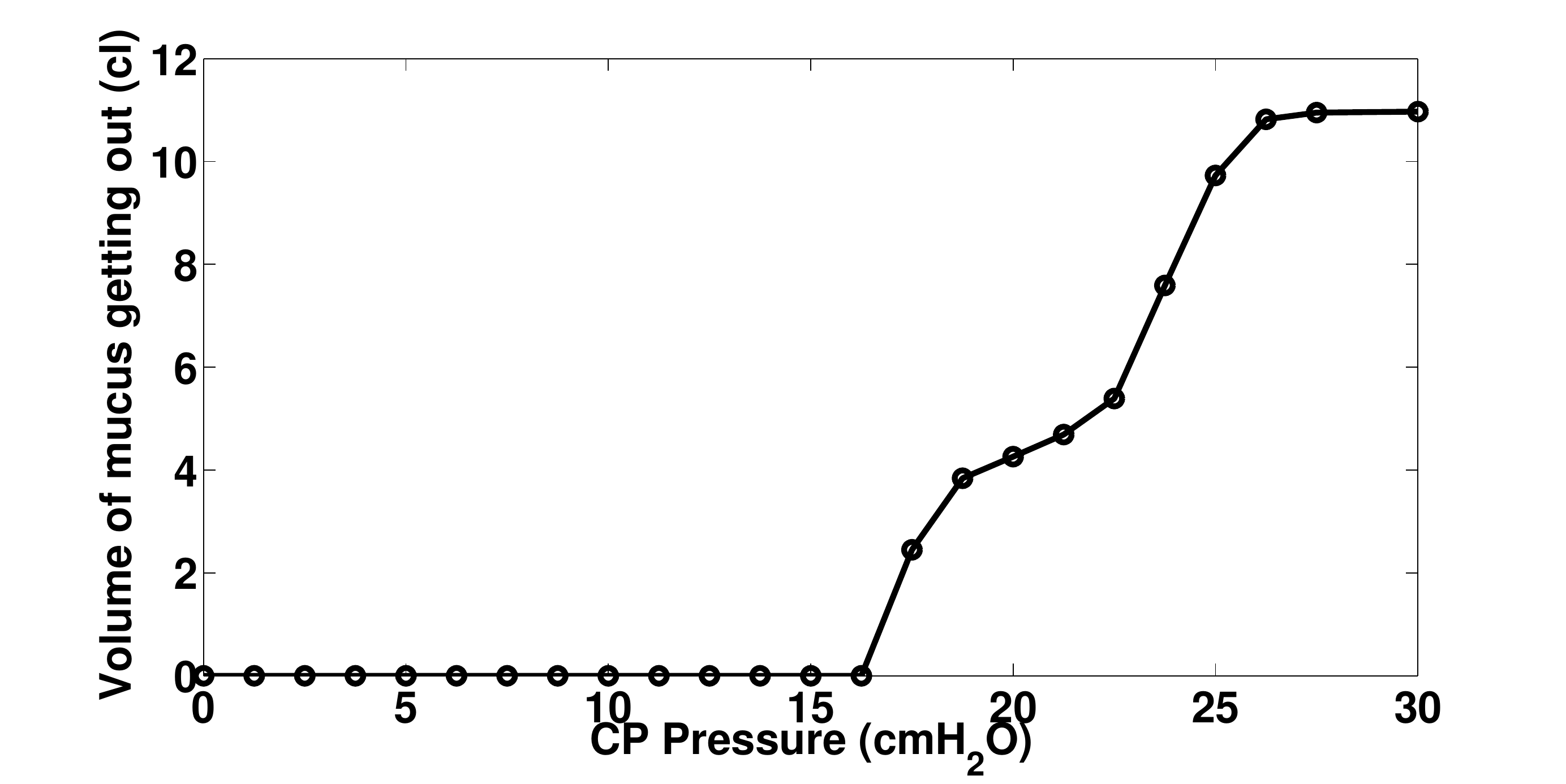}
B \includegraphics[height=2.5cm]{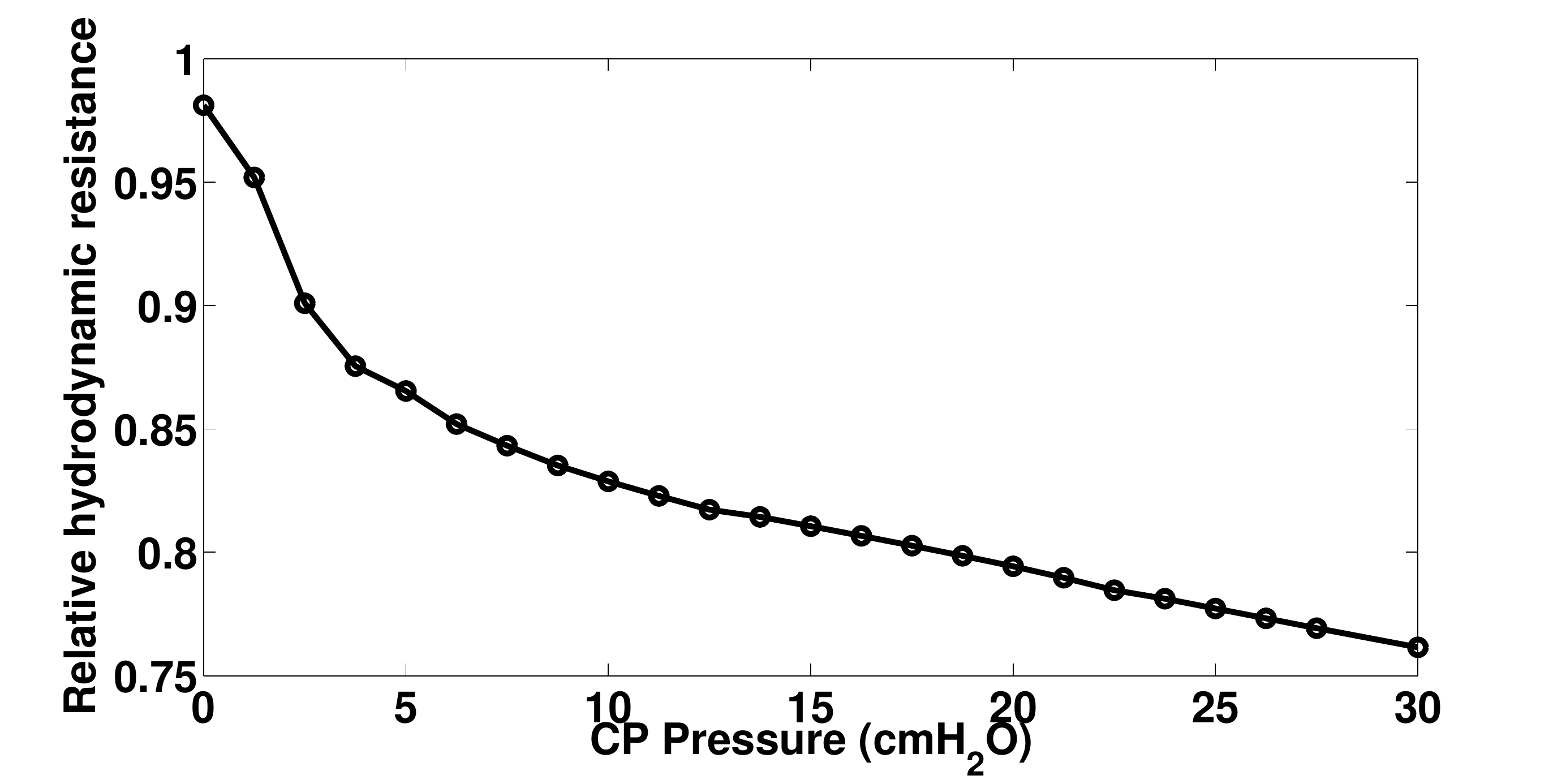}
C \includegraphics[height=2.5cm]{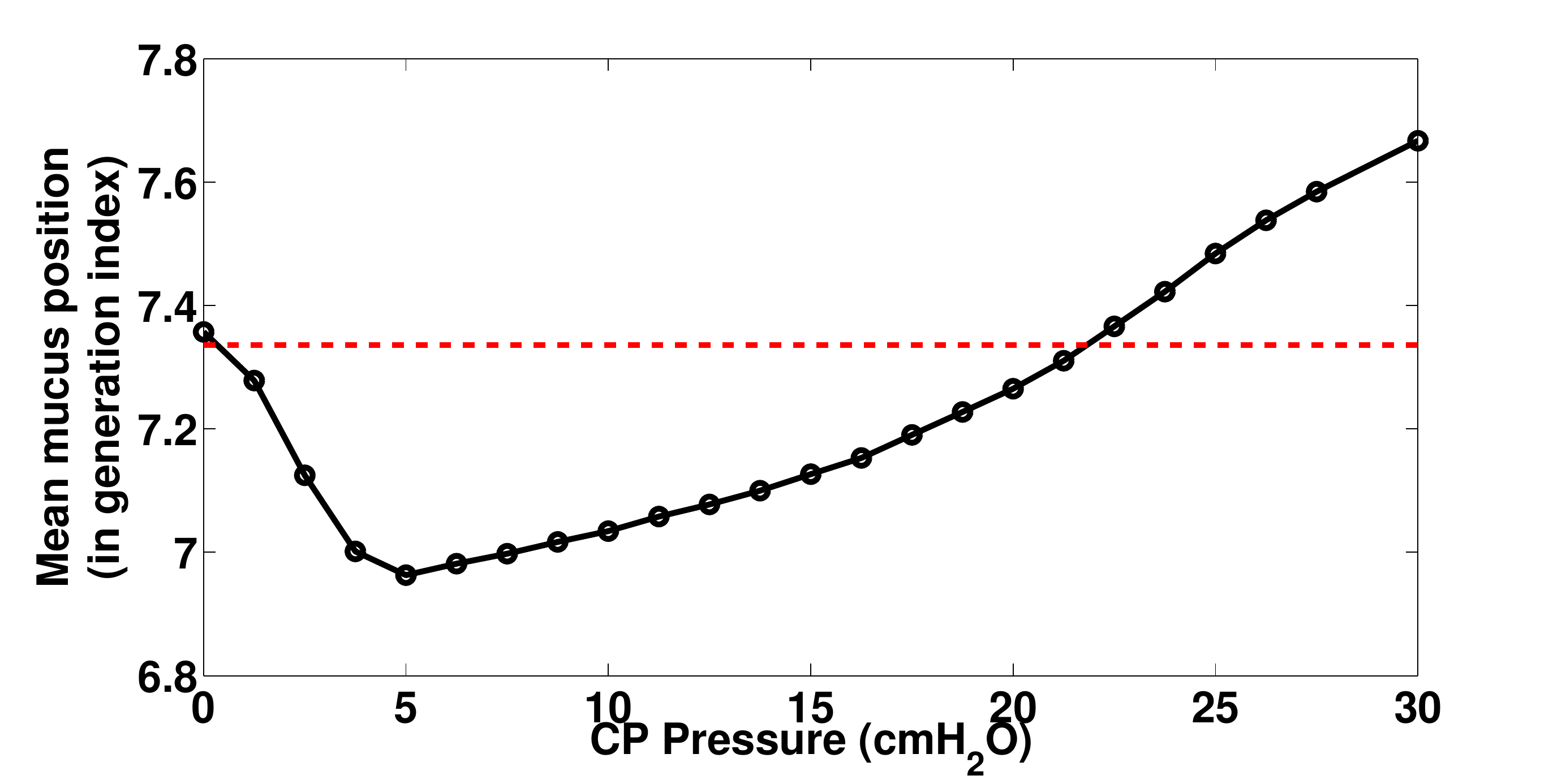}
\caption{A: Volume of secretions expelled at the end of the manipulation as a function of the pressure amplitude. B: Relative hydrodynamic resistance of the tree at the end of manipulation as a function of the pressure amplitude. C: Mean secretions position in the tree as a function of the pressure amplitude. Mean position is expressed in generation index.}
\label{rp2}
\end{figure}

We investigated the role of pressure $P_{cp}$ amplitude on mucus motion and distribution and on hydrodynamic resistance changes. Smaller chest pressures induce smaller air velocities and thus smaller shear stresses in the bronchi. As a consequence, the stress threshold that has to be overcome for secretions to be mobilized is only reached in highly obstructed bronchi. But the motion is quickly stopped when a new equilibrium between air flow and secretions distribution is reached. In particular, for chest pressures lower than $16.5 \ cmH_2O$, we do not observe in the model any secretions going out of the lung, see figure \ref{rp2}A. This does not mean however that secretions distribution has not been affected, as shown on figure \ref{rp2}B. An important point is that by performing the manipulation, secretions are moving in such a way that their distribution always decreases the total hydrodynamic resistance of the tree, down to a value corresponding to an equilibrium between secretions and air flows. The first steps of the manipulations are the more efficient, as shown on figure \ref{ManualPextRes}B. The higher the pressure, the lowest is the tree hydrodynamic resistance at the end of the manipulation. The first conclusion is then that the manipulations decrease the hydrodynamic resistance of the patient, which may lead to an improvement of patient breathing quality, even if no secretions get out of the tree.

\begin{figure}[h!]
\begin{tabular}{cc}
\includegraphics[height=2cm]{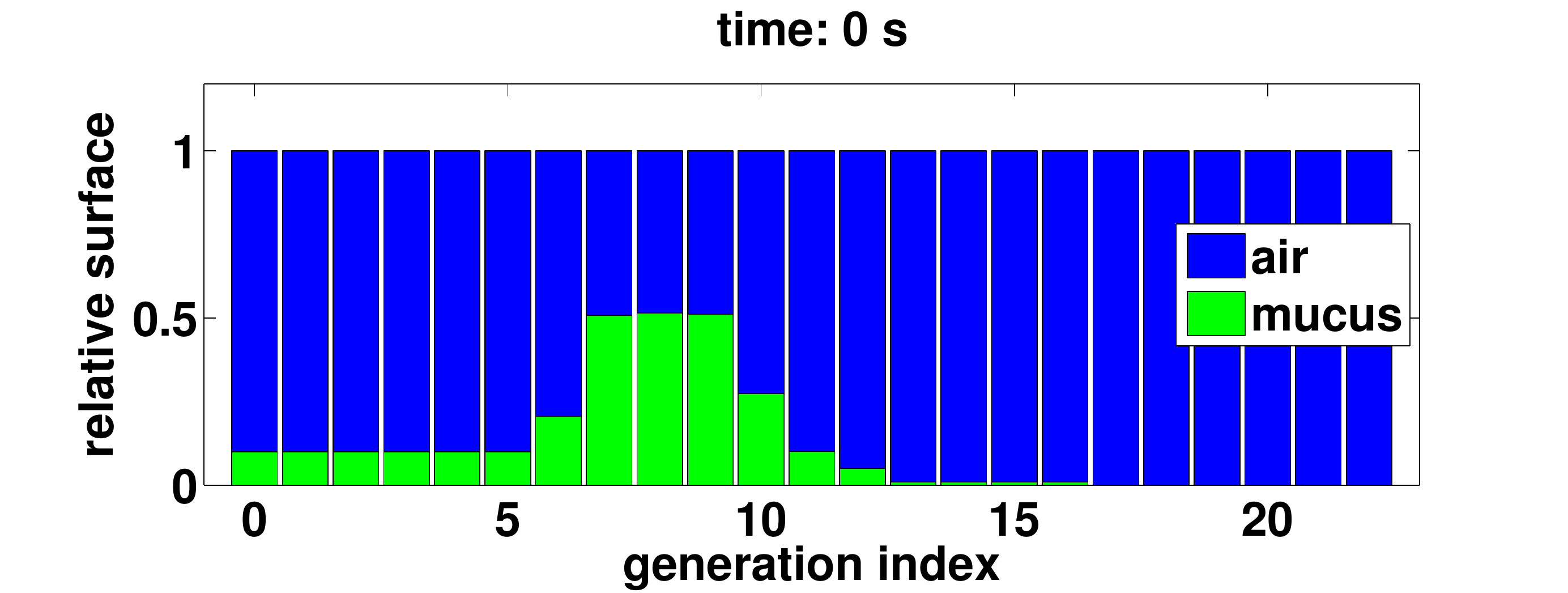}
& 
\includegraphics[height=2cm]{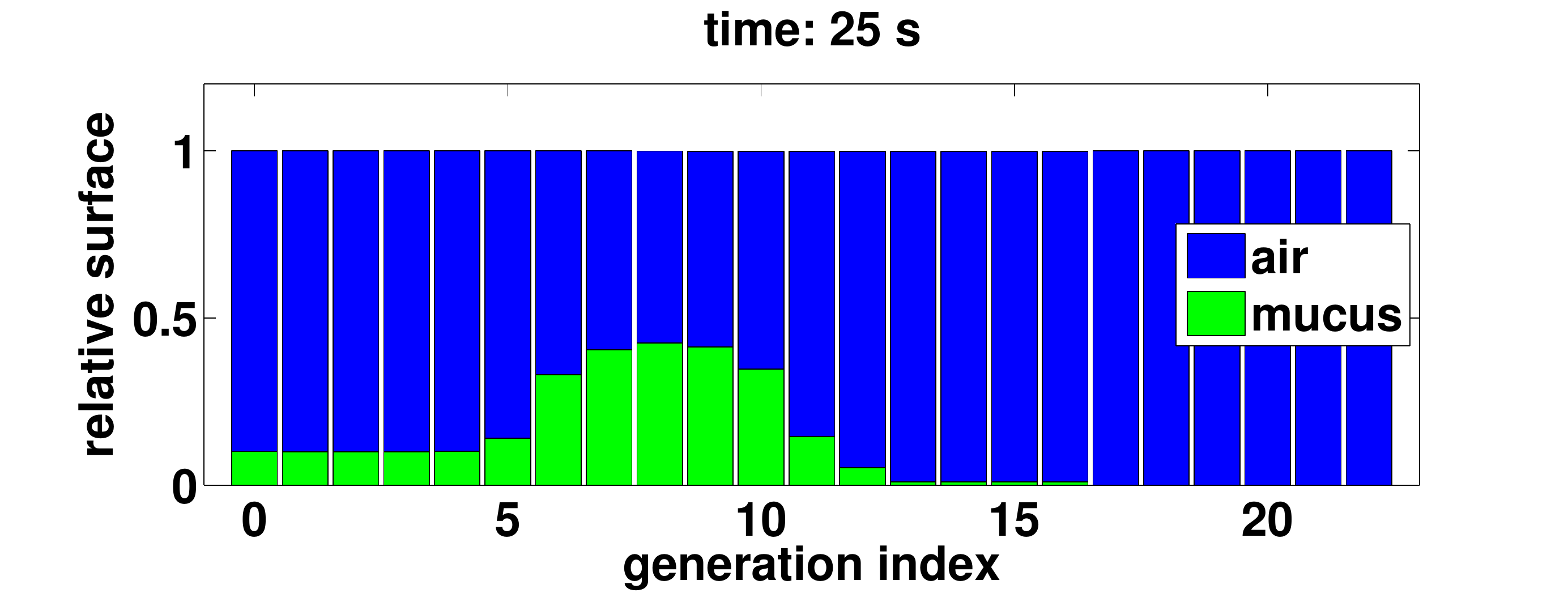}\\
\includegraphics[height=2cm]{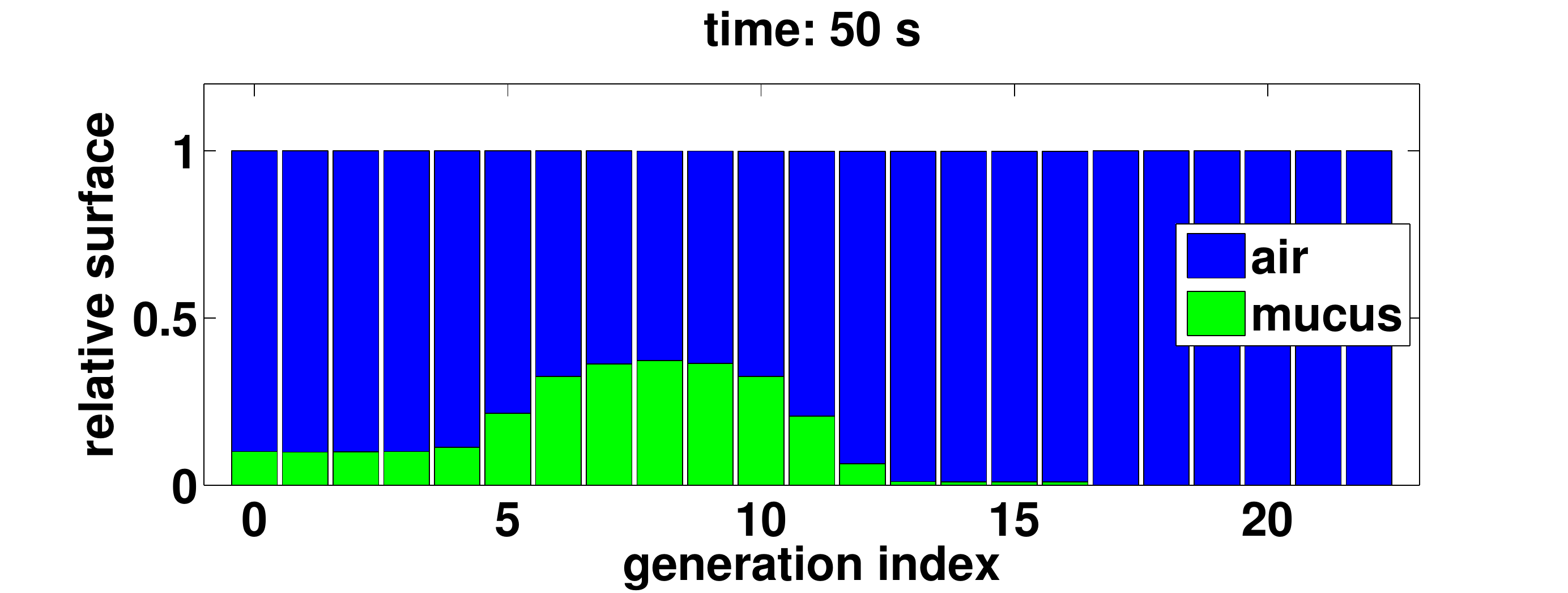}
& 
\includegraphics[height=2cm]{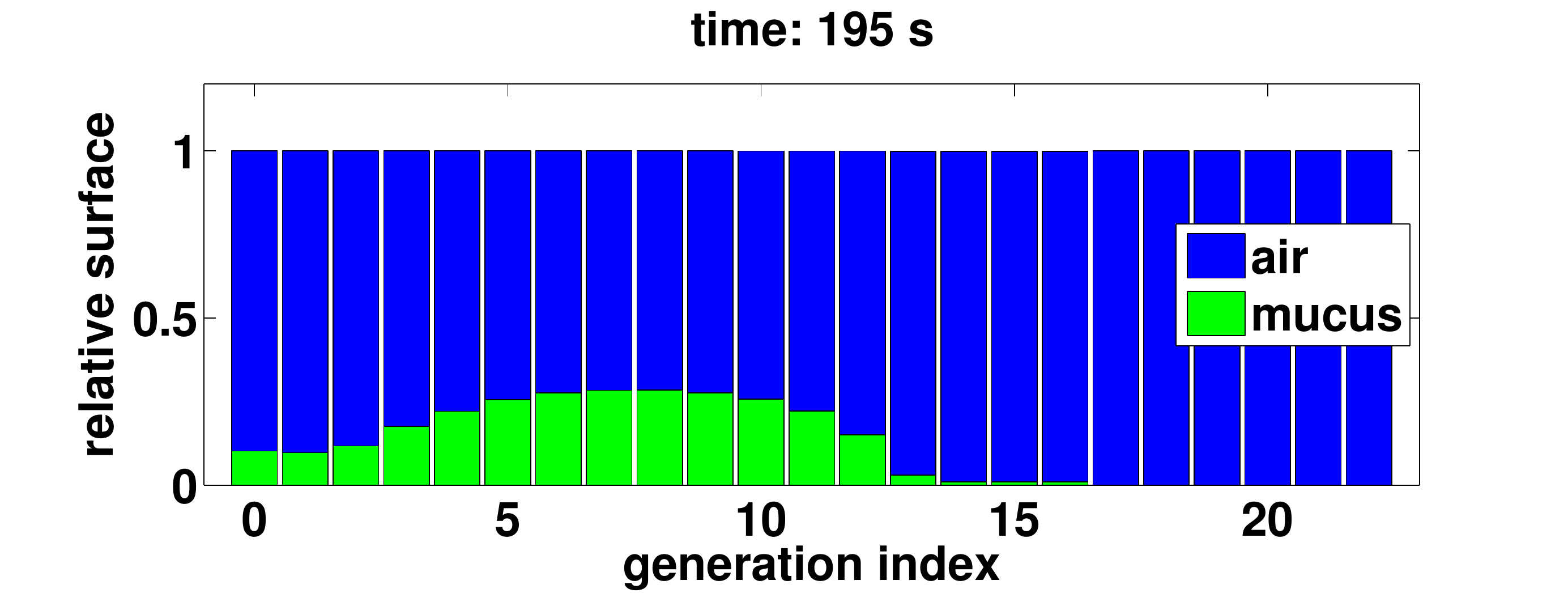}
\end{tabular} 
\caption{Evolution of mucus distribution in the bronchi during the manipulation, with $P_{cp} = 20 \ cmH_2O$. The blue bars correspond to the surface of bronchi lumen area relatively to bronchi initial surface. The green bars correspond to the surface fraction of bronchi lumen area obstructed by secretions. At the end of manipulation, secretions have been globally motioned upward the tree, however some secretions have been pushed deeper in the lung, see generation $11-13$.}
\label{rp2MucDist}
\end{figure} 

Secretions distribution in the tree are motioned from one generation to the next, and because manipulation pressures are applied during expiration, mucus tends to go up the tree, at least for moderate pressures amplitude, see figure \ref{rp2}C. The mechanism is as follow: during pressure application, secretions are moving upward in the tree from one generation to the next, and thus fill the upper generations. Lung's recoil from the manipulation and lung opening due to inspiration are correlated to an inlet air flow in the lung, which potentially moves the secretions back into the lower generations. This backward secretions motion depends on how much secretions have accumulated during the pressure application. If the pressure is high, then secretions accumulate a lot in the upper bronchi, reducing drastically air lumen area, and by thus they are more affected by the inward air flow. This is why for high pressures, although there is less secretions in the tree at then end, they are globally deeper after the manipulation than before the manipulation. This phenomenon can be seen on figure \ref{rp2MucDist} where generations $11$ to $13$ have more secretions at the end of the manipulation.

\subsection{Mechanical physiotherapy}
\label{mechaPhysio}

\begin{figure}[h!]
A
\includegraphics[height=3cm]{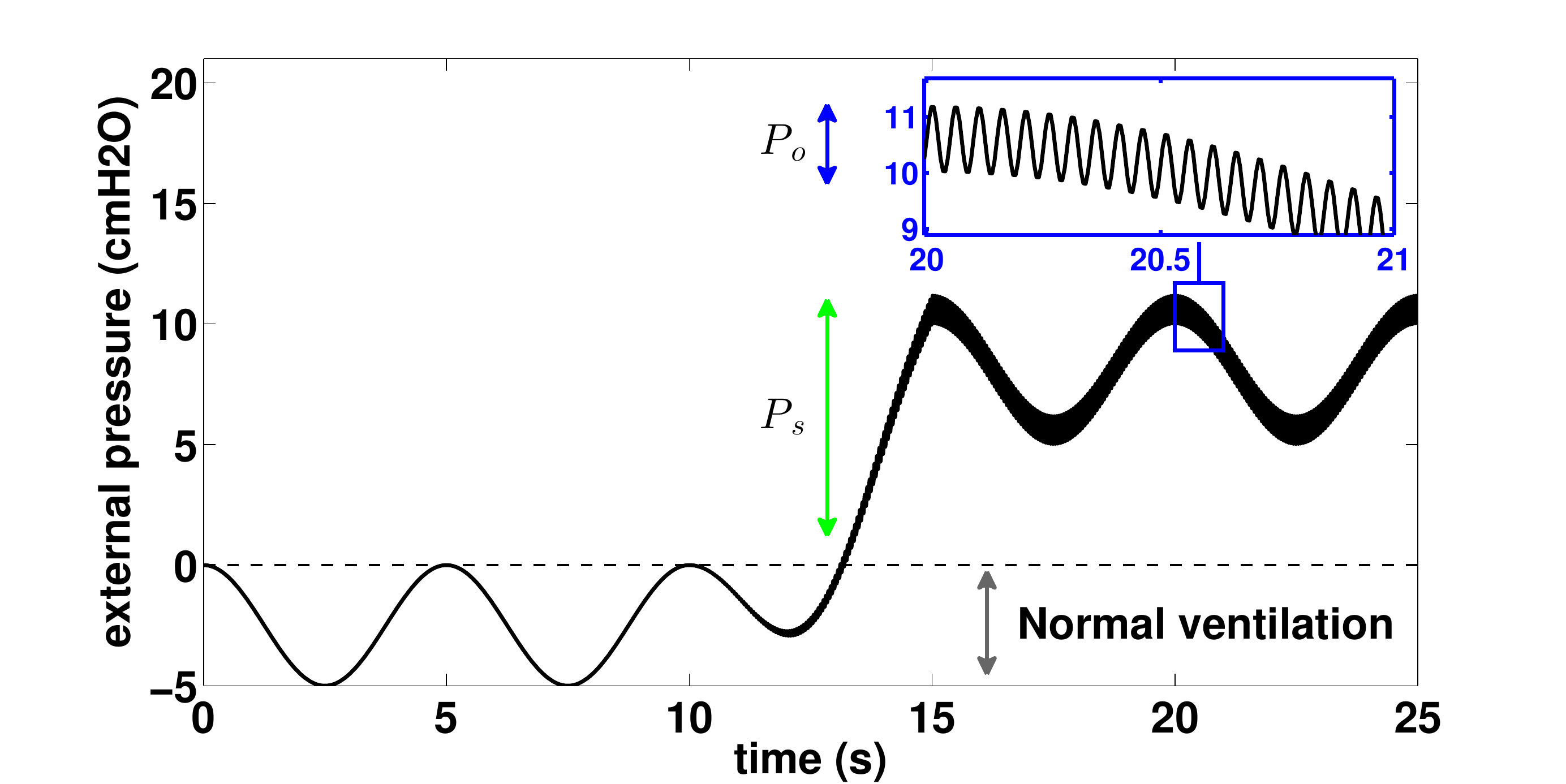}
B
\includegraphics[height=3cm]{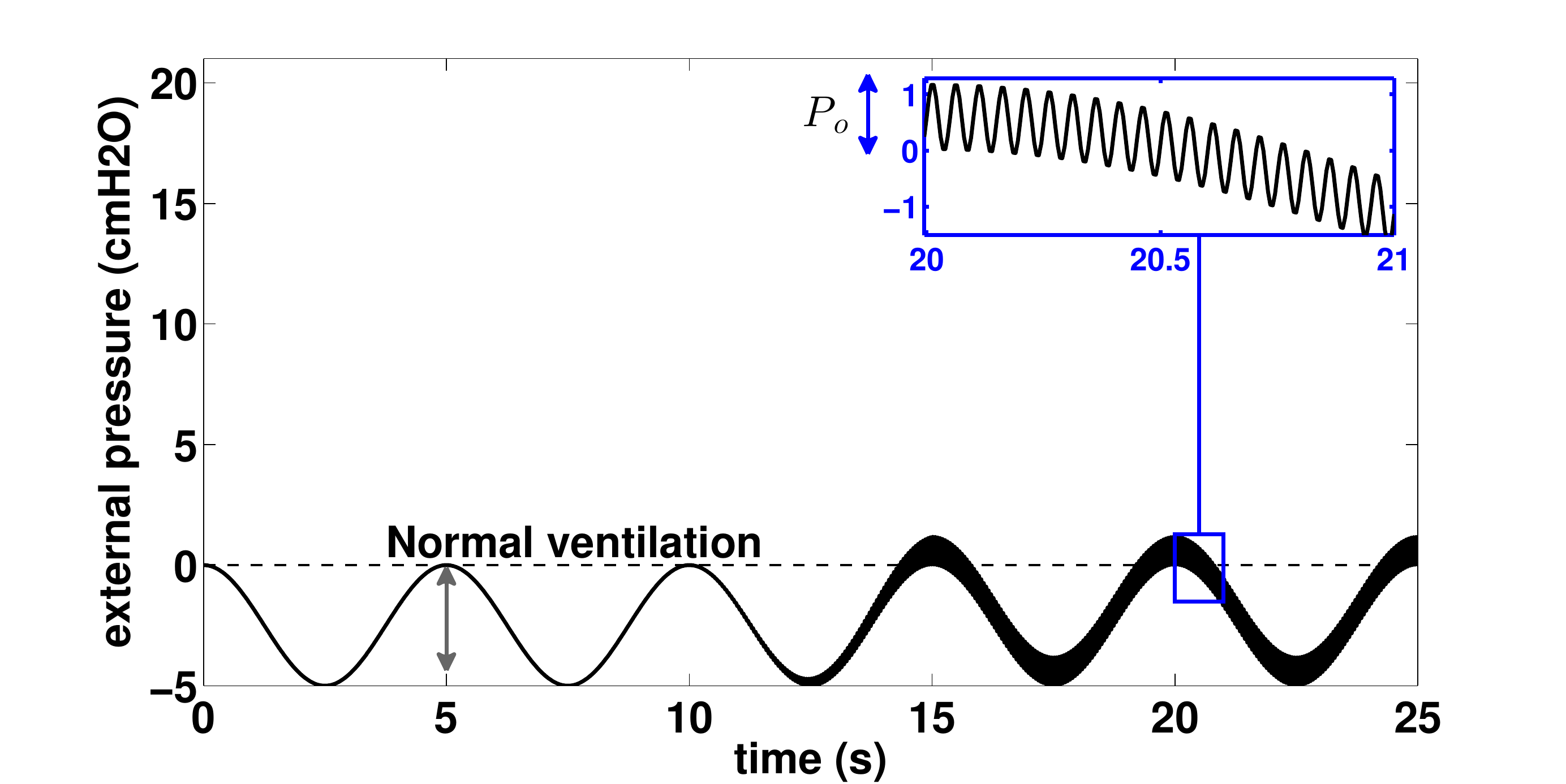}
\caption{Chest pressure used as an input for our model to mimic mechanical chest physiotherapy based on high frequency chest wall oscillations ($20 \ Hz$ in this example). Pressure is assumed homogeneous on the thorax. A: Pressure distribution on the thorax to mimic devices using chest compression (CC) using a static pressure. B: Pressure distribution on the thorax to mimic devices using focused pulses technique (FPT) without static pressure ($P_s = 0$).}
\label{Pext_hfcwo}
\end{figure}

There exists a whole range of mechanical chest physiotherapy devices that are able to help secretions expectoration \cite{hristara-papadopoulou_current_2008, arens_comparison_1994, mitchell_hfcwo_2013}. We will focus on devices that work by applying oscillatory positive pressures on the chest, a method that is referred to as {\it high frequency chest wall oscillation} (HFCWO). HFCWO covers different ways to apply the pressure on the thorax. {\it Chest compression} (CC) devices \cite{arens_comparison_1994} apply a global static pressure on the chest, and low amplitude oscillations are applied on the whole thorax, thus working at low lung volume. {\it Focused pulses technique} (FPT) devices \cite{mitchell_hfcwo_2013} do not apply a global static pressure but instead apply high amplitude oscillations on specific locations on the chest using pistons. In this section, we propose to mimic with our model the effects of these two types of devices. We chose an input external pressure that adds to normal ventilation a static pressure $P_s$ and an oscillatory pressure with amplitude $P_o$ and frequency $f$, see figure (\ref{Pext_hfcwo}):
$$
P_{ext}^{oscill}(t) = P_{ext}^{ventil}(t) + P_{s} + \frac{P_o}2 \sin( 2 \pi f t)
$$

In the case of focused pulses technique, it is necessary to account for the pistons covering only a small fraction of the thorax. Thus, we use equivalent pressures $P_o$ as inputs of the model. They represent the typical values applied by the pistons weighted by the surfaces ratio between pistons' heads and thorax. As for manual chest physiotherapy modeling, manipulation pressures are applied for one session of $230$ seconds. The $10$ first and last seconds are normal ventilation. The time range $10 s$ to $15 s$ is used to start smoothly the manipulation by applying progressively the manipulation pressure. The time range $215 s$ to $220 s$ is used to stop smoothly the manipulation by decreasing progressively the manipulation pressure.

\subsubsection{An example of high frequency chest wall oscillation}

In this section, we study the predictions of the model with an input mimicking the effects of HFCWO with a static pressure $P_s = 5.6 \ cmH_2O$ and an oscillatory pressure $P_o = 1.2 \ cmH_2O$ at $f = 20 \ Hz$. Because of the static pressure, both ventilation and pressures oscillations are applied on a lung with low volume. Most particularly, this affects the net bronchi lumen area and the pleural pressure value and response \cite{agostoni_static_2011, dangelo_statics_2005}.

\begin{figure}[h!]
A
\begin{minipage}[l]{0.6\textwidth}
\begin{tabular}{cc}
\includegraphics[height=2cm]{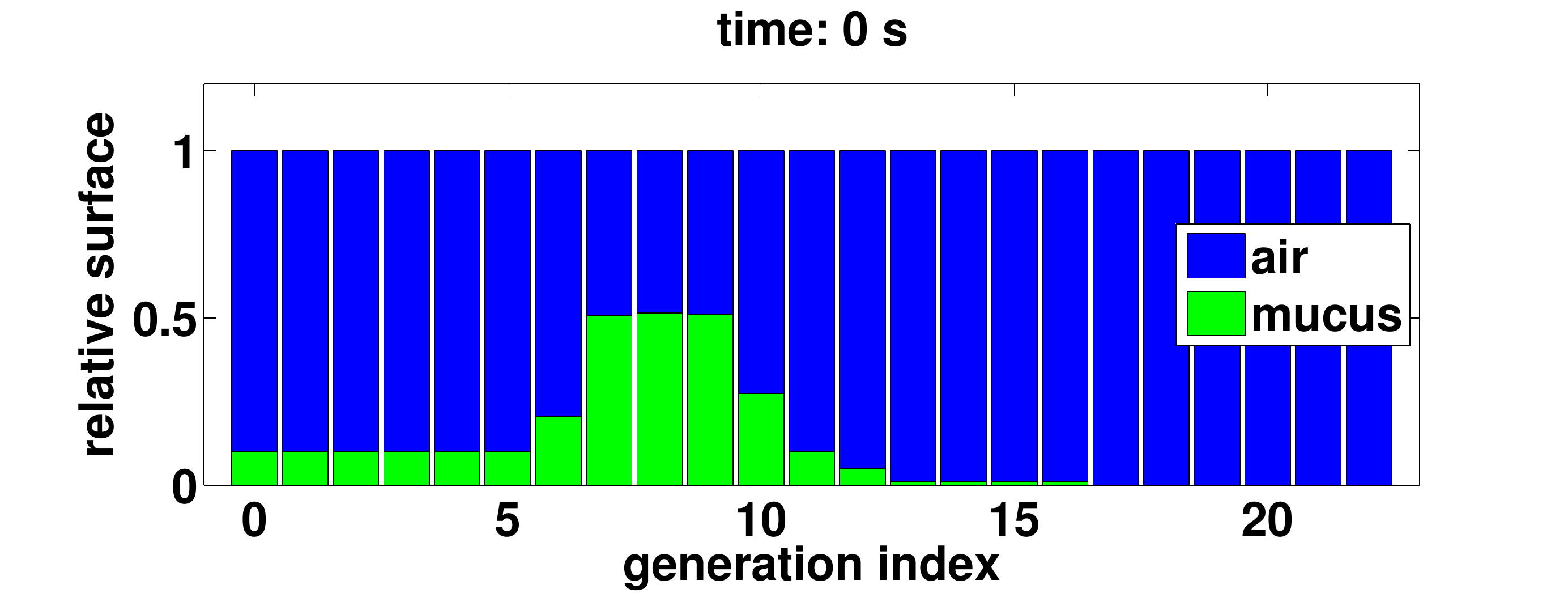}
& 
\includegraphics[height=2cm]{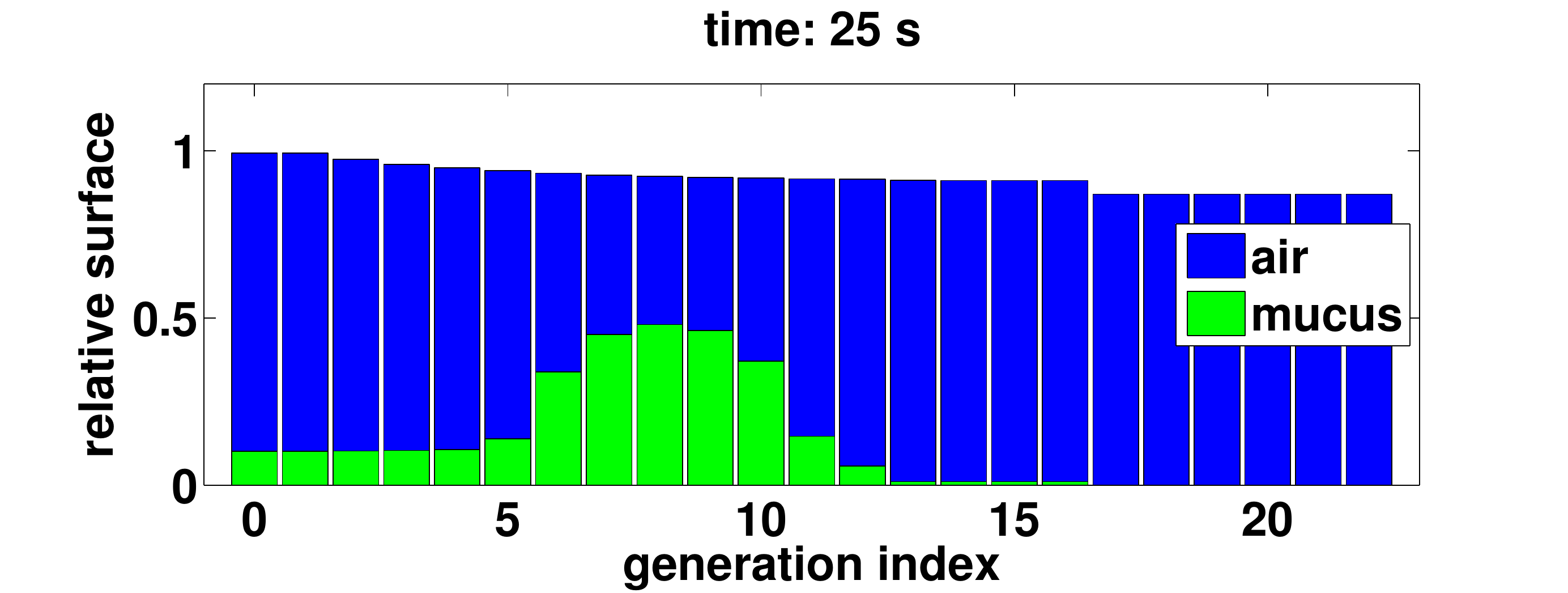}\\
\includegraphics[height=2cm]{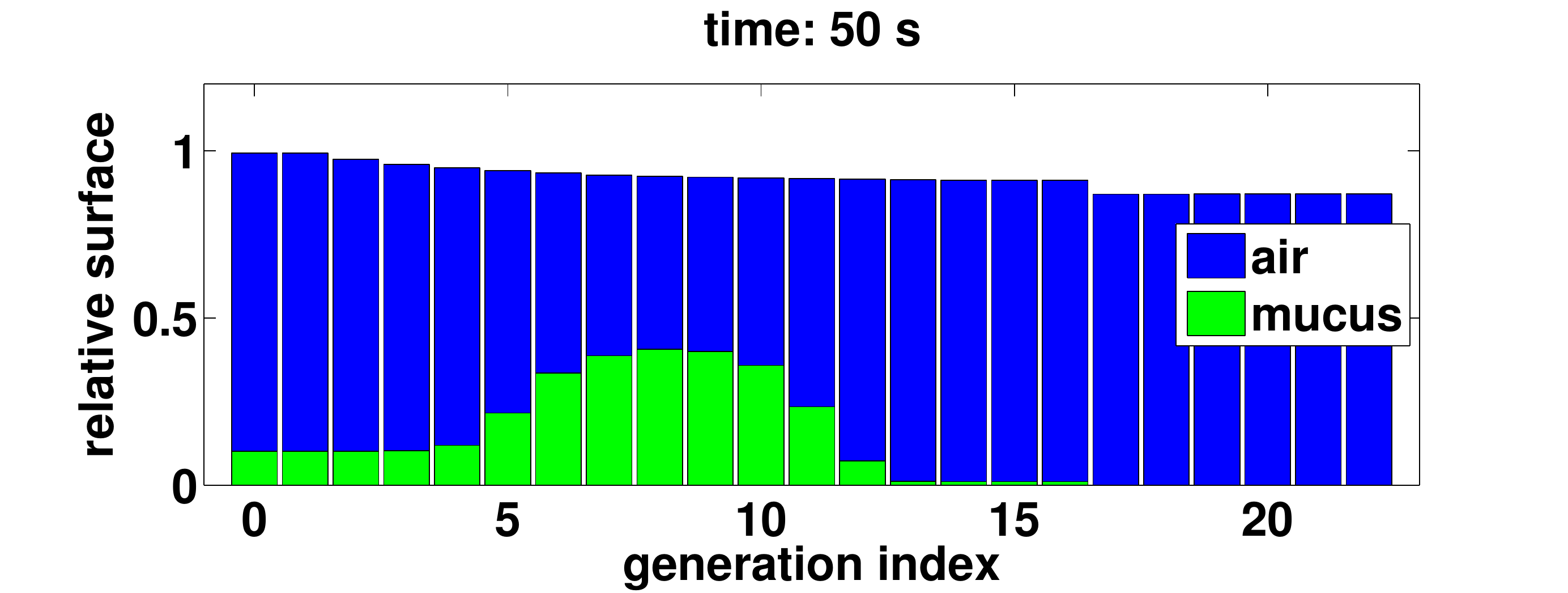}
& 
\includegraphics[height=2cm]{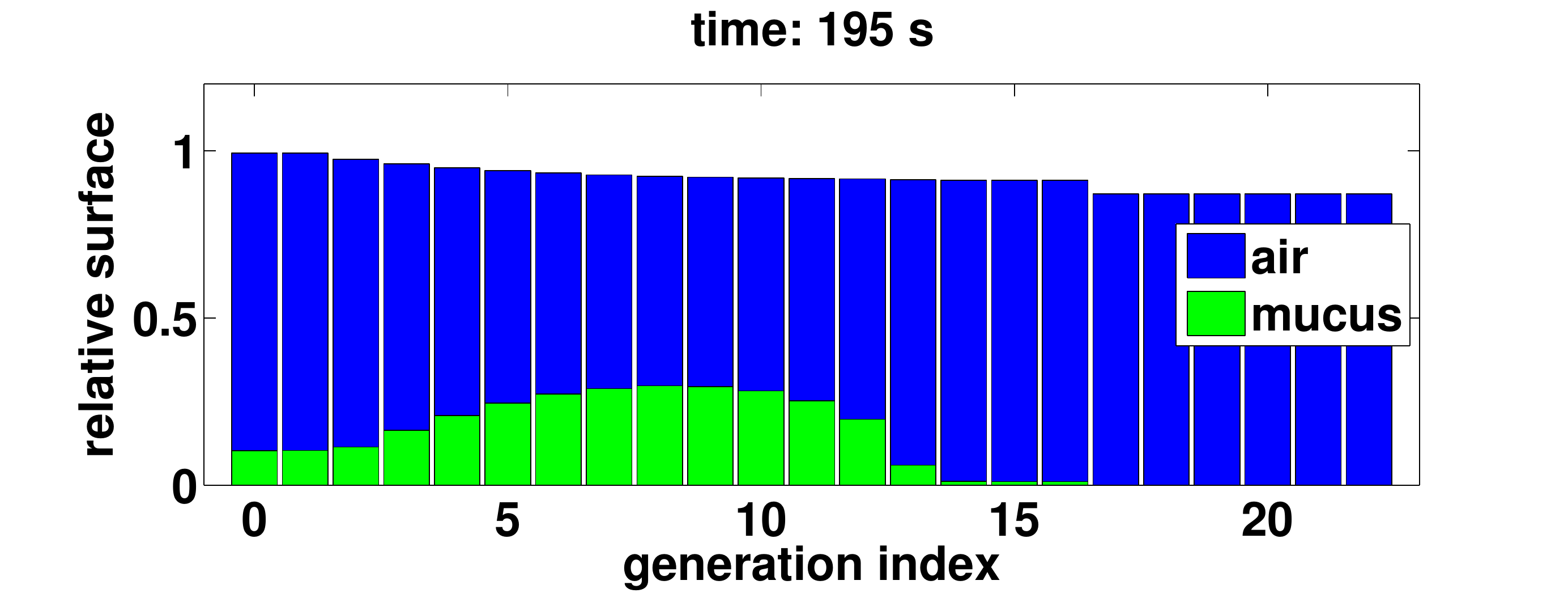}
\end{tabular}
\end{minipage}
B
\begin{minipage}[l]{0.3\textwidth}
\includegraphics[height=3cm]{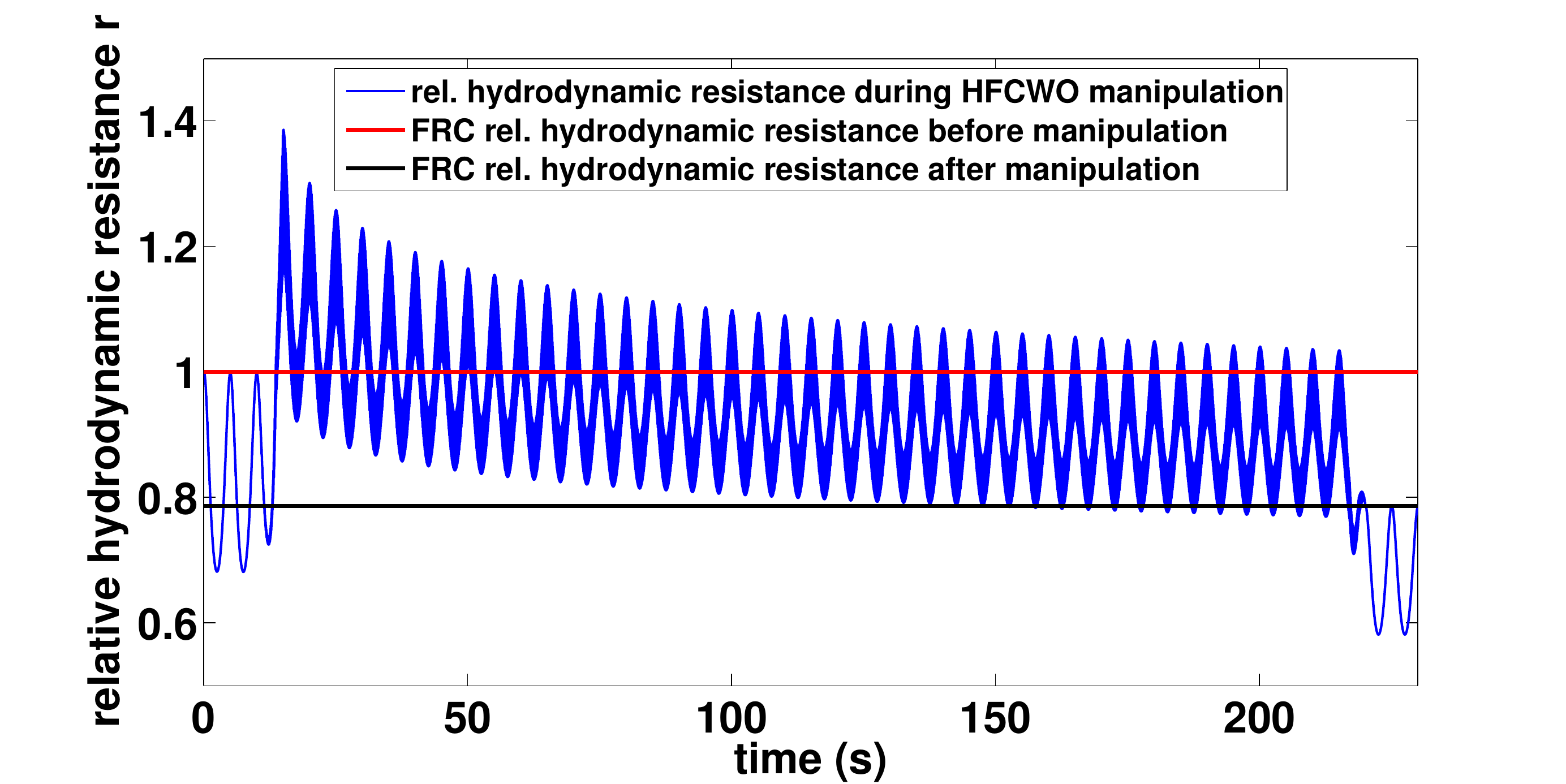}
\end{minipage}
\caption{{\bf A}: Evolution of mucus distribution in the bronchi during the manipulation with $P_s = 5.6 \ cmH_2O$ and $P_o = 1.2 \ cmH_2O$. {\bf B}: Relative hydrodynamic resistance changes during a manipulation with $P_{cp} = 20 \ cmH2O$ (blue). The red line corresponds to FRC relative hydrodynamic resistance before the manipulation ($r_{|t=0s}=1.00$) while the black line corresponds to FRC relative hydrodynamic resistance at the end of the manipulation ($r_{|t=230s}=0.79$).}
\label{hfcwoEx}
\end{figure} 
As for manual chest physiotherapy, our model predicts that the manipulation redistributes the secretions in the tree in such a way it reduces the hydrodynamic resistance of the tree by about twenty percents at the end of the manipulation, see figure \ref{hfcwoEx}A. The manipulation efficiency is higher at the beginning and decreases with time, as shown on figure \ref{hfcwoEx}B. Because the static pressure and the oscillations occurred during the whole ventilation cycle, the secretions spread a little more down the tree than for manual chest physiotherapy. Initially, mean mucus generation in the tree is $7.34$, and it reaches $7.47$ at the end of the manipulation, see figure \ref{hfcwoPs}. With the parameters used in this section, our model predicts that no secretions are going out of the tree.

\subsubsection{Role of static pressure $P_s$}

\begin{figure}[h!]
A \includegraphics[height=2.5cm]{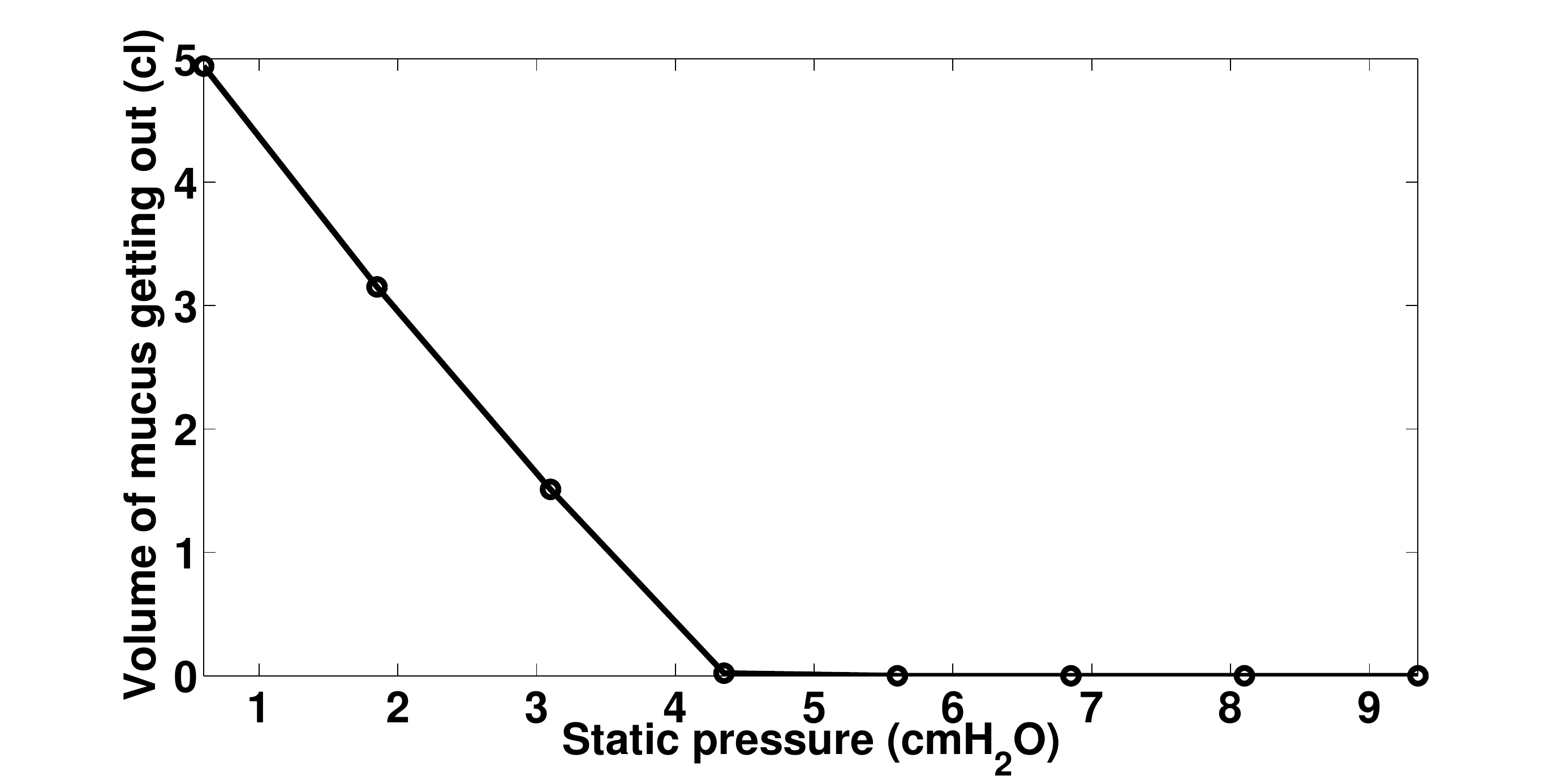}
B \includegraphics[height=2.5cm]{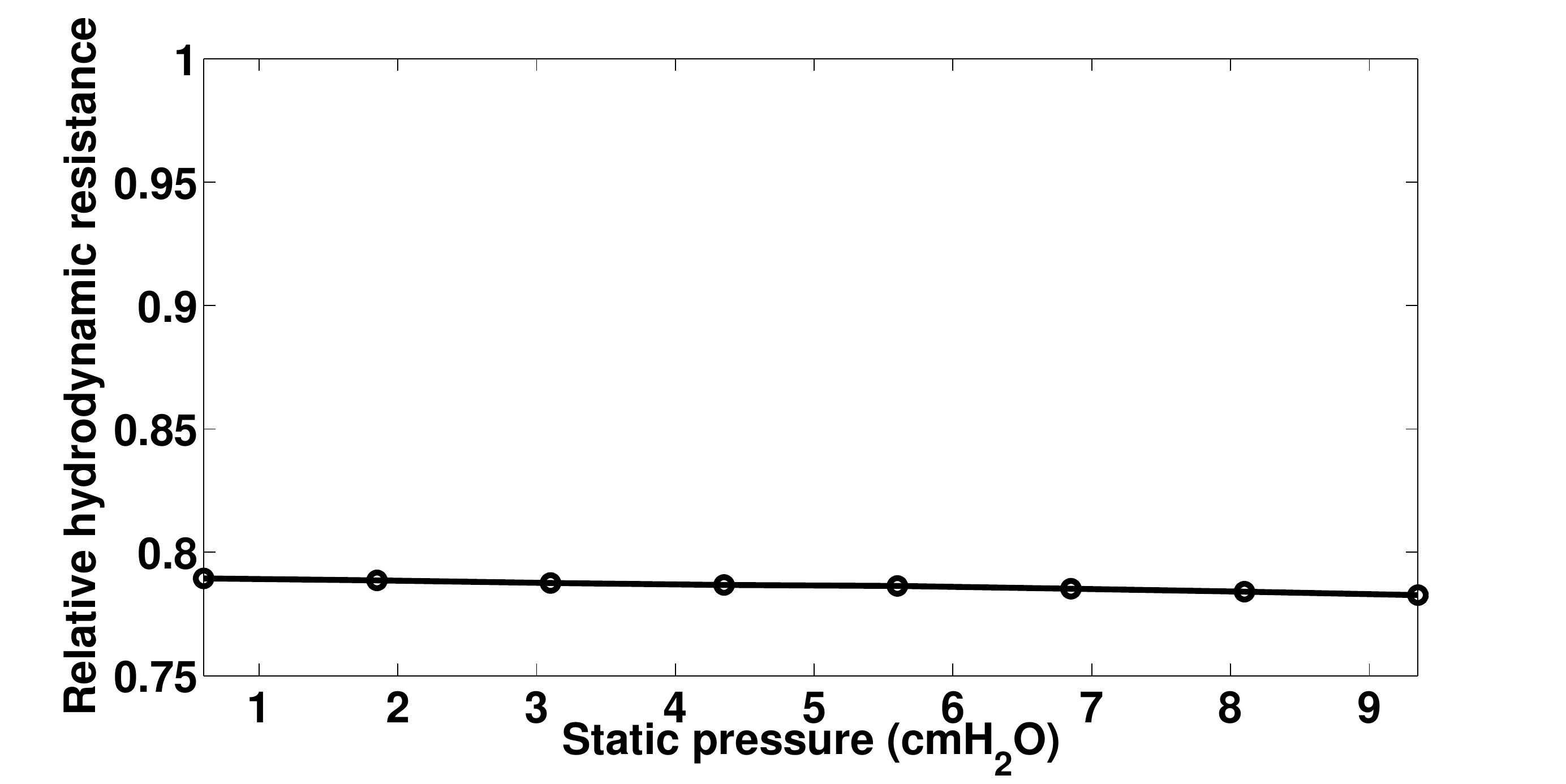}
C \includegraphics[height=2.5cm]{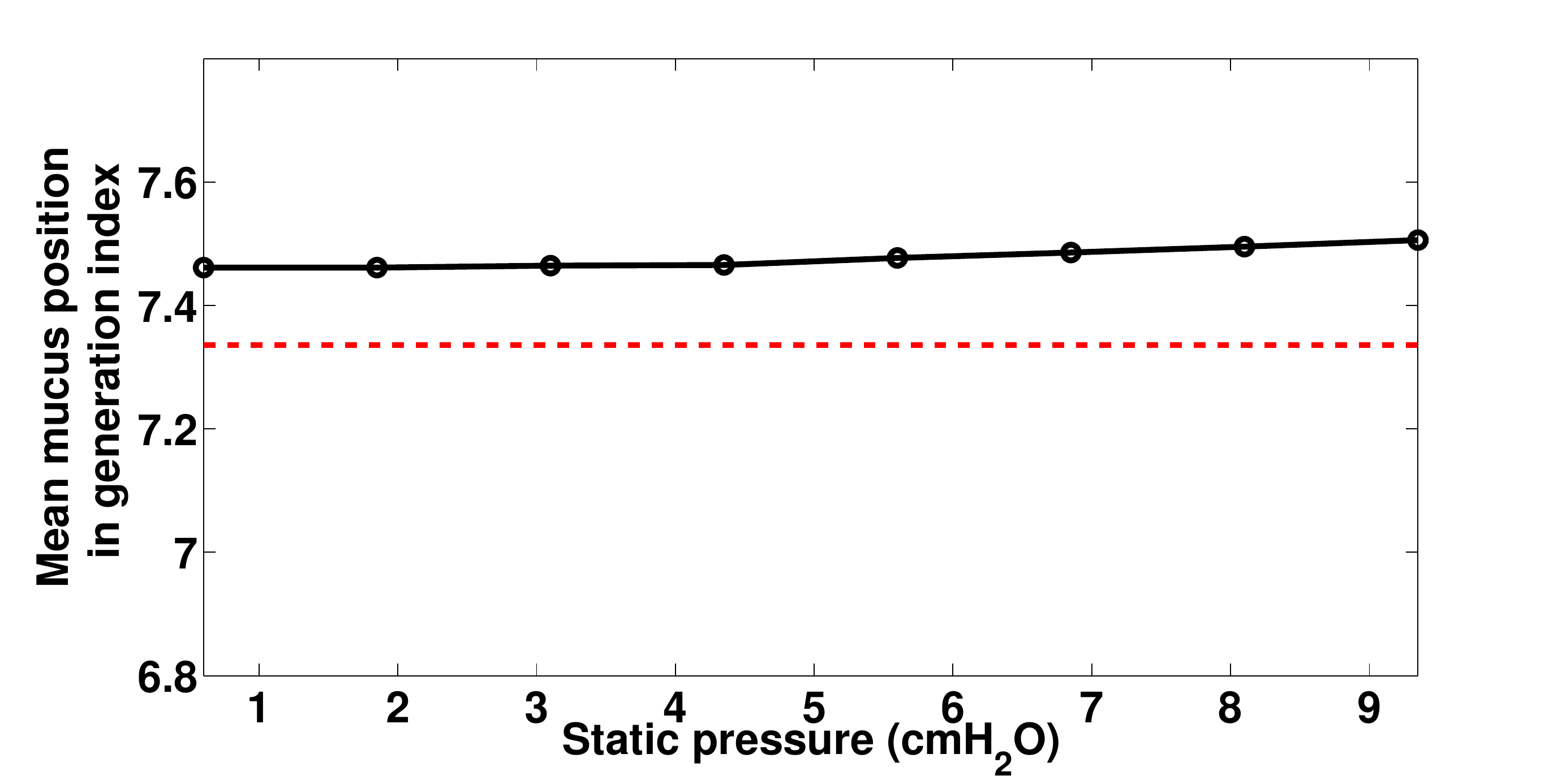}
\caption{Role of static pressure $P_s$, with $P_o = 1.2 \ cmH_2O$ and $f = 20 \ Hz$, data computed at the end of the manipulations for time $t=230 \ s$. A: Volume of mucus getting out of the tree. B: Relative hydrodynamic resistance of the tree. C: Mean mucus position in the tree, in generation index. The red dashed line represents the value before the manipulation.}
\label{hfcwoPs}
\end{figure}

We investigated the role of the static pressure $P_s$ that controls the lung volume at which the manipulation is performed. Oscillating pressure $P_o$ was fixed to $1.2 \ cmH_2O$ with a frequency $f$ at $20 \ Hz$. Surprisingly, we found that static pressure has only small effects on the final hydrodynamic resistance and on the mucus spreading in the tree. Increasing the static pressure leads to a slight improvement of the hydrodynamic resistance at the end of the manipulation, see figure \ref{hfcwoPs}B. Larger static pressure also sends secretions a bit deeper in the tree, and increases slightly the mean mucus generation at the end of the manipulation, see figure \ref{hfcwoPs}C. Finally, increasing static pressure reduces to amount of secretions getting out of the tree, see figure \ref{hfcwoPs}A. 

These results indicate that, in the frame of our model, increasing the static pressure may have no real effect. Indeed, increasing the static pressure goes with a constriction of the bronchi, which increases the hydrodynamic resistance of the bronchial tree, as shown in the previous example on figure \ref{hfcwoEx}B. The air flow created by the oscillating pressure is then smaller if the static pressure is larger. Shear forces determine the motion of the secretions, and in a bronchi, they are proportional to the ratio between the air flow rate in the bronchi and the diameter of the bronchi at the power three. Our results suggest that the increase on shear stress gained by smaller airways diameters is almost compensated by the loss in term of air flow rate.

\subsubsection{Influence of the oscillating pressure amplitude $P_o$ and frequency $f$}

Air flow in a bronchi determines the shear stress applied on the secretions and thus their motion. Air flow in a bronchi is partially related to velocity of the bronchi volume changes, see equation (\ref{fm1}). Thus secretions motion is correlated to bronchi volume change velocity. This quantity is actually the ratio between the amount of volume change over the time during which this volume change occurs. The amount of volume change is given by oscillating pressure amplitude, while the time for these changes to occur is given by the oscillating pressure frequency. Thus, these two quantities are ought to play an important role on manipulation efficiency. 

\begin{figure}[h!]
A \includegraphics[height=2.5cm]{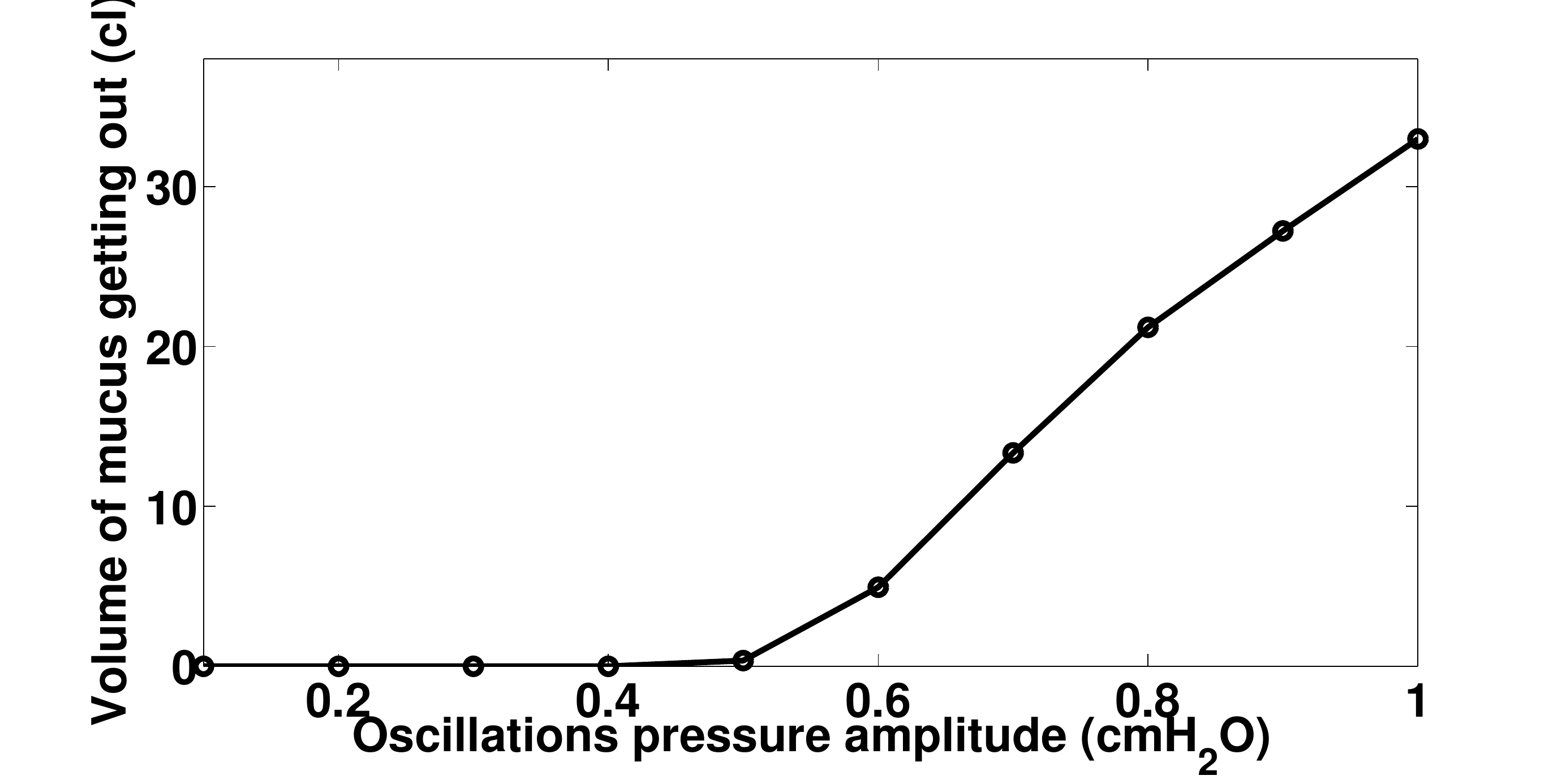}
B \includegraphics[height=2.5cm]{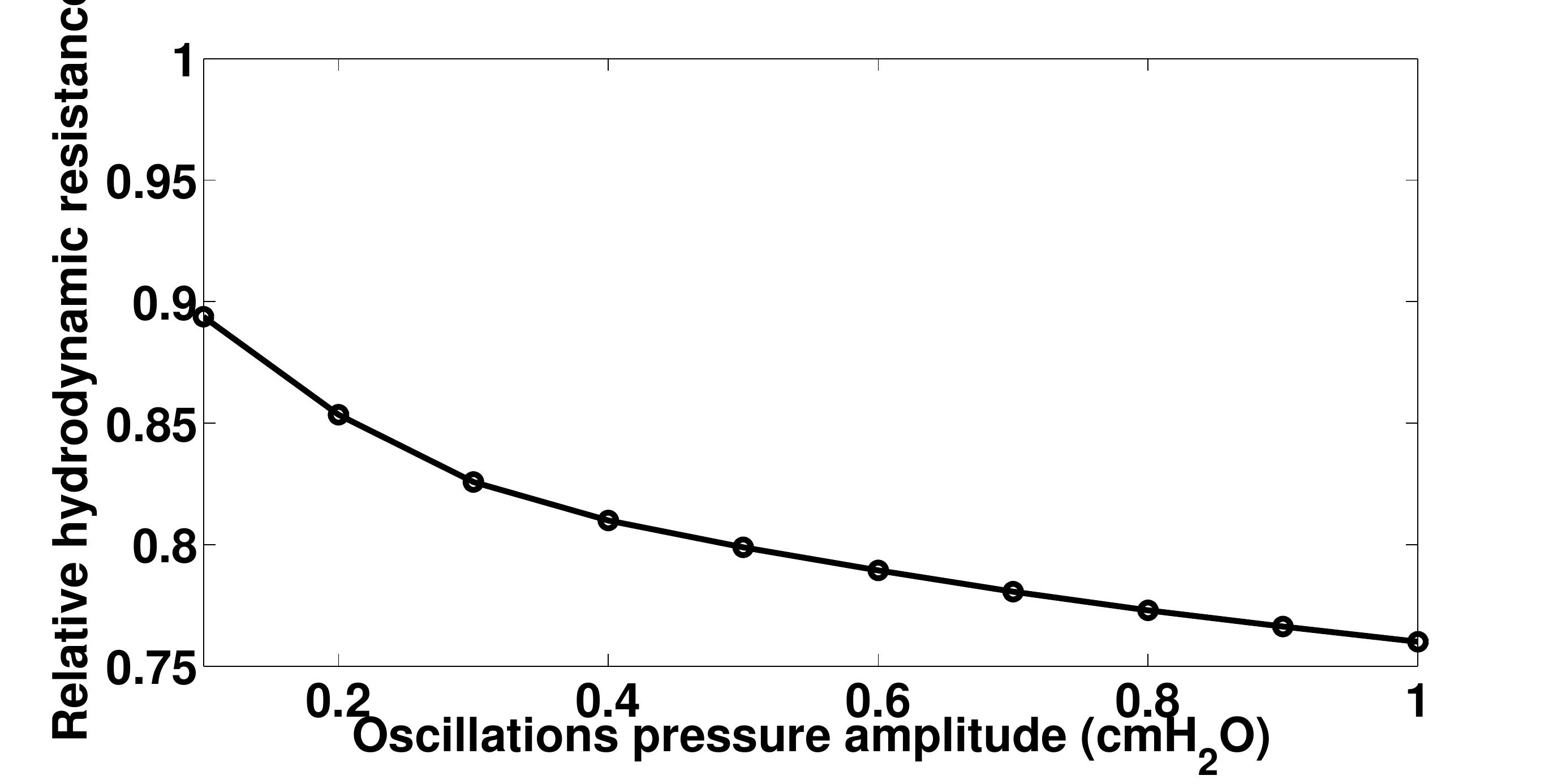}
C \includegraphics[height=2.5cm]{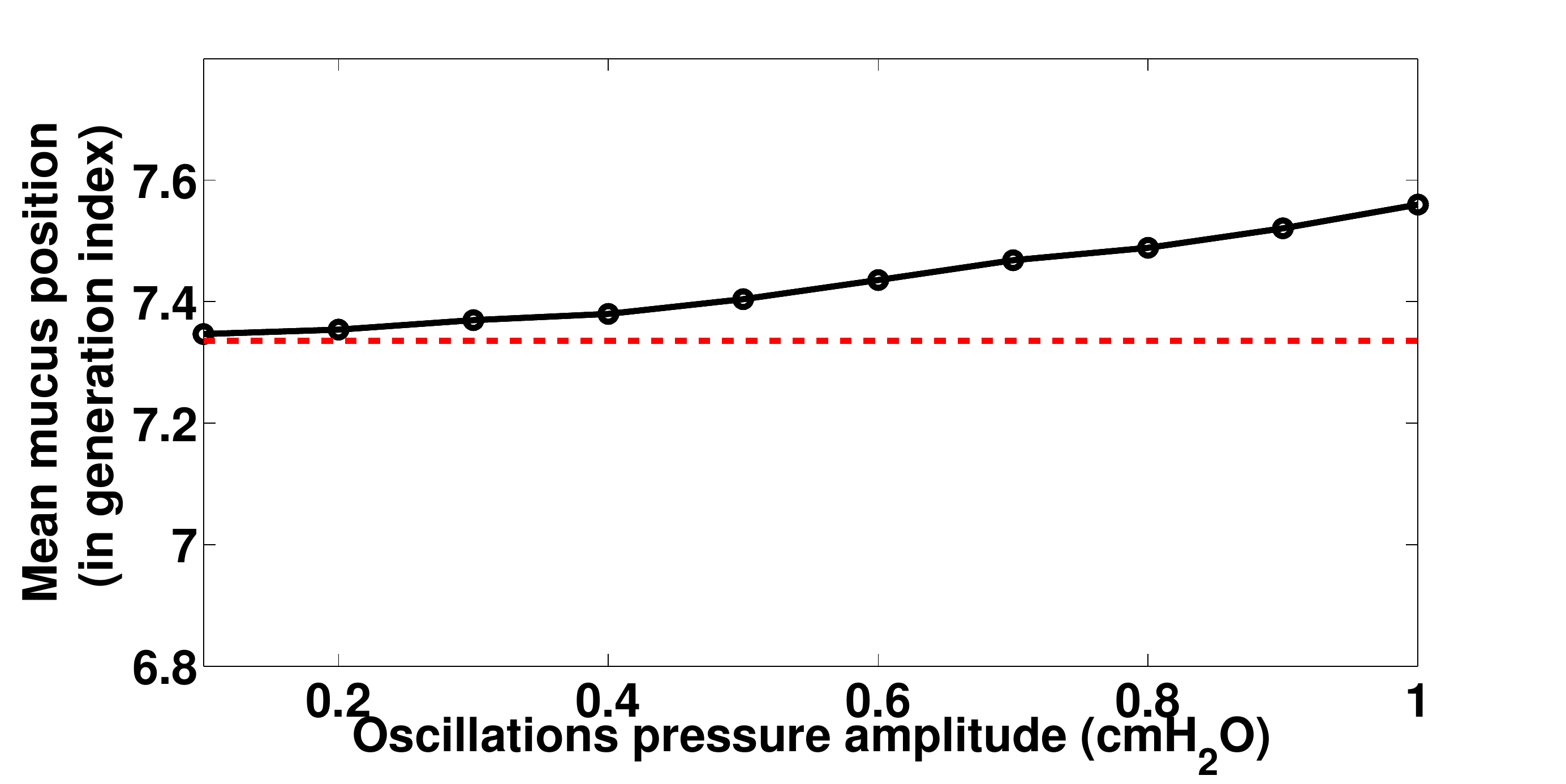}
\caption{Role of oscillating pressure $P_o$, with $P_s = 0.6 \ cmH_2O$ and $f = 20 \ Hz$, data computed at the end of the manipulations for time $t=230 \ s$. A: Volume of mucus getting out of the tree. B: Relative hydrodynamic resistance of the tree. C: Mean mucus position in the tree, in generation index. The red dashed line represents the value before the manipulation.}
\label{hfcwoPo}
\end{figure}

We investigated first the role of the oscillating pressure $P_o$. Because static pressure has only a small effect in our model, we assume $P_s = 0.6 \ cmH_2O$ which makes the total pressure applied on the thorax oscillate between $0$ and $1.2 \ cmH_2O$. The frequency $f$ is kept at $20 \ Hz$. As predicted, the larger the amplitude of the oscillations, the more efficient is the manipulation, see figure \ref{hfcwoPo}. As before, a more efficient mechanical manipulation tends to send more mucus down the tree, as shown on figure \ref{hfcwoPo}C, because oscillations are symmetric and applied during the whole cycle of ventilation.

\begin{figure}[h!]
A \includegraphics[height=2.5cm]{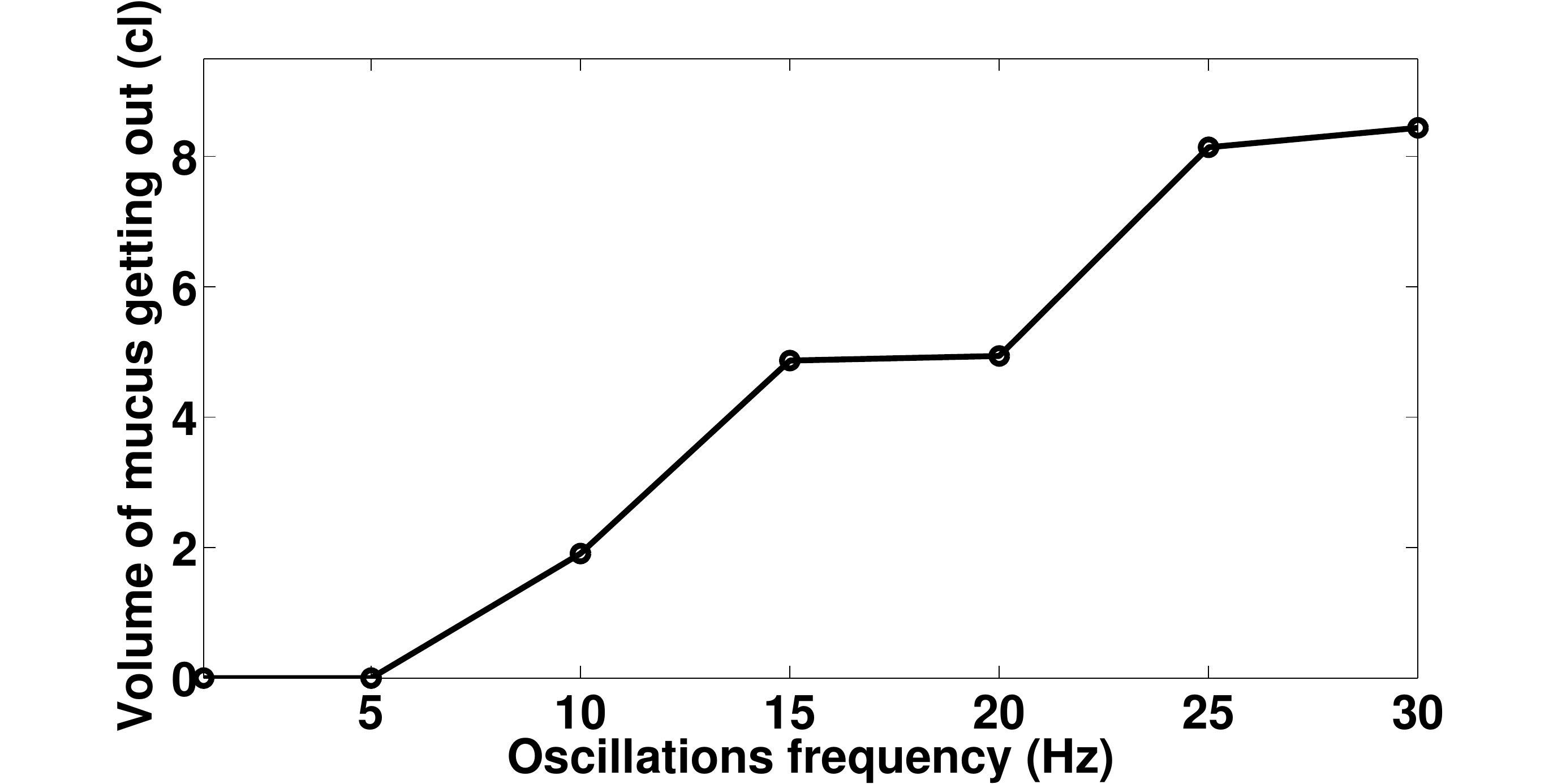}
B \includegraphics[height=2.5cm]{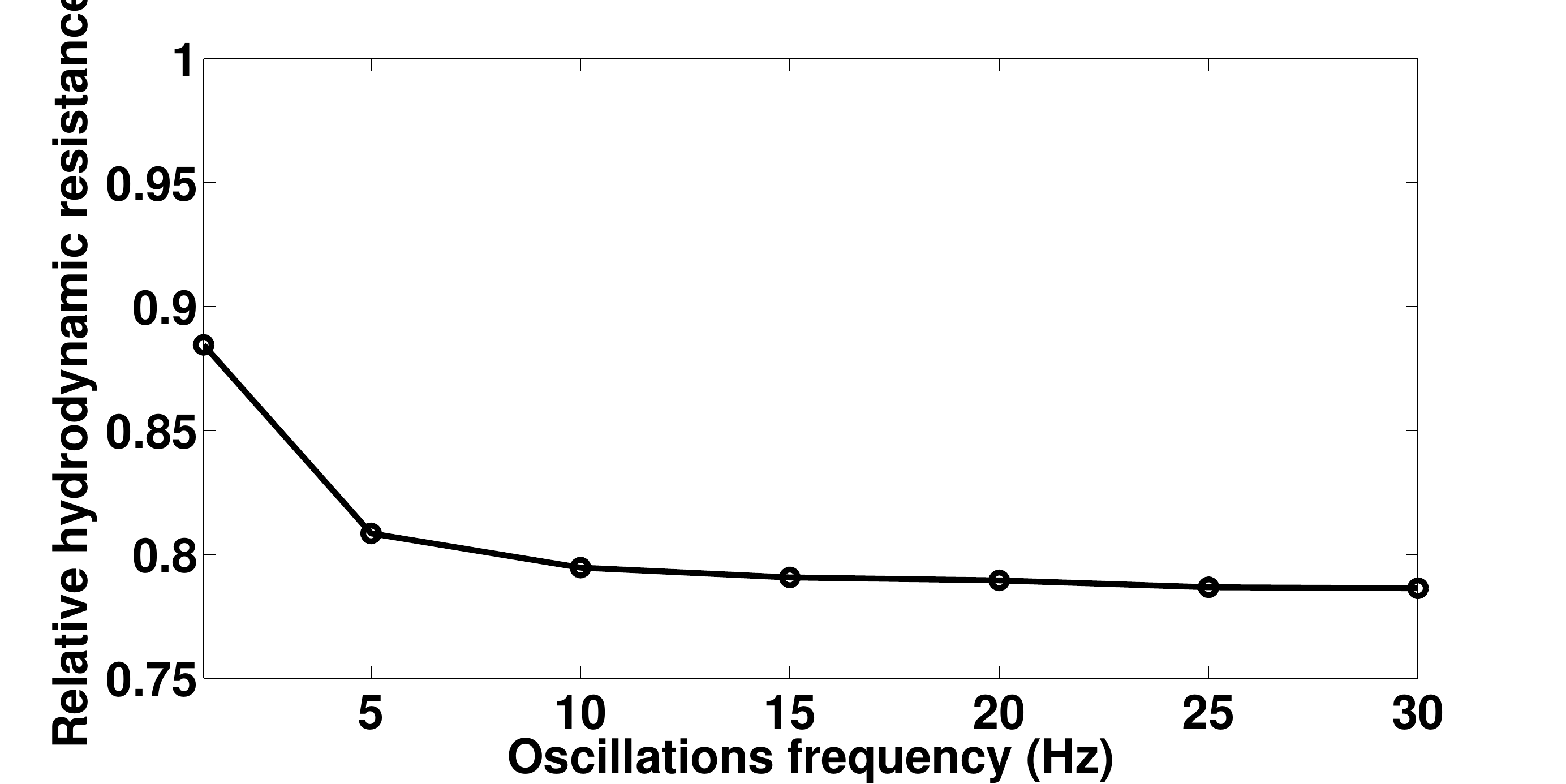}
C \includegraphics[height=2.5cm]{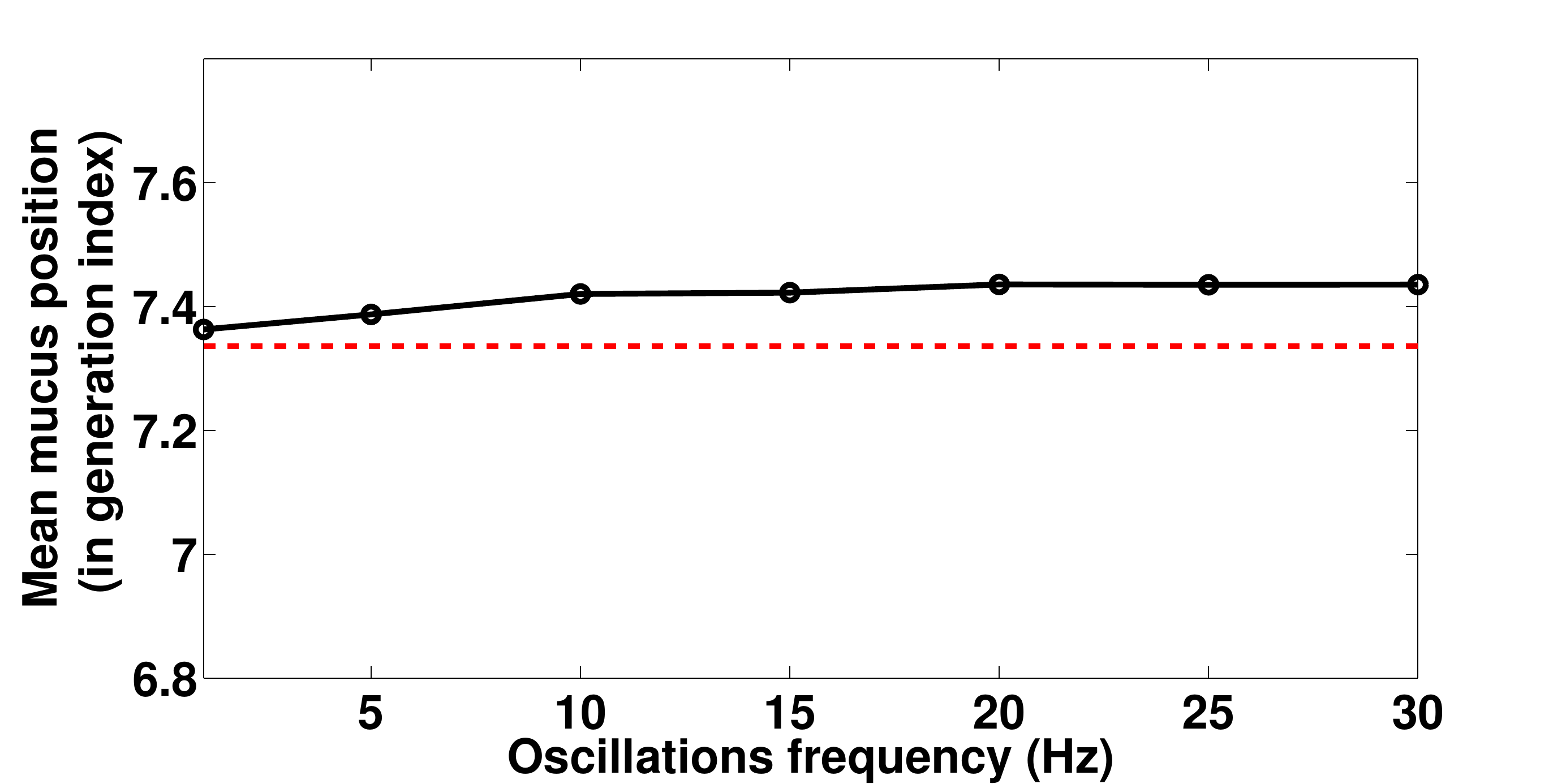}
\caption{Role of oscillating pressure frequency $f$, with $P_s = 0.6 \ cmH_2O$ and $P_o = 1.2 \ cmH_2O$, data computed at the end of the manipulations for time $t=230 \ s$. A: Volume of mucus getting out of the tree. B: Relative hydrodynamic resistance of the tree. C: Mean mucus position in the tree, in generation index. The red dashed line represents the value before the manipulation.}
\label{hfcwoFreq}
\end{figure}

The frequency $f$ also plays an important role on the manipulation efficiency. There is a minimal frequency under which the manipulations have low efficiency. Once this minimal frequency is reached, the hydrodynamic resistance and the mean mucus position in the tree are only slightly affected by any supplementary increase of the frequency, as shown on figure \ref{hfcwoFreq}. Only the quantity of secretions that get out of the tree continue to increase regularly with the frequency, as shown on figure \ref{hfcwoFreq}A, mainly because the secretions in the root of the tree (mimicking the trachea) are all the more mobilized than the frequency is high.

\subsection{Shrek number: a measure for chest physiotherapy efficiency}
\label{shrek}

The mucus behavior in a generation $i$ can be predicted with the ratio between the air constraints on the mucus measured by the air shear stress $\tau_i$ in an airway and the shear stress threshold $\sigma_0$ of the mucus. In an airway with radius $r_i$ and length $l_i$, the air shear stress is $\tau_i = 4 \mu_a \Phi_a^{(i)} / (\pi r_i^3)$. We define the instantaneous Shrek number $Sh$ as the average of the ratios $\tau_i / \sigma_0$ over the tree generations:
\begin{equation}
\label{shrekNumberDef}
\text{Instantaneous Shrek\footnote{Named after the main character of the Dreamworks animation {\it Shrek} (2001). Shrek use his own secretions as core material for building everyday objects.} number: } iSh = \frac1N \sum_{i=0}^{N-1} \frac{4 \mu_a \Phi_a^{(i)}}{\pi r_i^3 \sigma_0}
\end{equation}
The instantaneous Shrek number can be approximated with a formula easier to evaluate, using lung hydrodynamic resistance $R_{aw}$ and air flow at mouth level $\Phi_a$, see details in appendix \ref{approxShrek}. This approximation is based on the hypothesis that the ratios of length over diameter of all the airways are equal to $3$, which corresponds to the mean value measured \cite{tawhai_ct-based_2004}:
\begin{equation}
\label{shrekNumber}
iSh \sim \frac{R_{aw} \Phi_a}{12 N \sigma_0}
\end{equation}
The larger the instantaneous Shrek number, the more secretions are mobilized. An instantaneous Shrek number smaller than $1$ does not mean that mucus is not mobilized, since Shrek number represents an average over the generations. It means however that in many generations, the mucus is probably not mobilized. However it is important to notice that airways impaired by mucus will have the highest Shrek numbers in the tree. Because chest physiotherapists are modulating the flow in their manipulation to keep mucus moving in the tree, they are indirectly and intuitively modulating the instantaneous Shrek number to keep it high enough to get a response from mucus. These modulations are paired with the real-time changes of lung hydrodynamic resistance. 

Finally, the instantaneous Shrek number can be reformulated as a function of the homogeneous external pressure used in our model. The new formulation involves quantities related to lung function and to manipulation: the compliance of the lung $C_L$ that relates the external pressure to the lung volume change, and the velocity with which changes in external pressure occurs $\Delta P_{ext} / \Delta t$:
\begin{equation}
\label{shrekNumber2}
iSh \sim \frac{C_L R_{aw}}{12 N \sigma_0} \frac{\Delta P_{ext}}{\Delta t}
\end{equation}
\noindent This formula comes from the simple fact that air flow in the lung equals lung volume change $\Delta V_L$ divided by the duration for this change $\Delta t$. Also the compliance of the lung checks $C_L = \Delta V_L / \Delta P_{ext}$. Combining these equations leads to the alternative formulation for instantaneous Shrek number $iSh$ in equation (\ref{shrekNumber2}).

To be able to compare different chest physiotherapy manipulations in the hypotheses of our model, we compute the Shrek number which is the time average of the instantaneous Shrek number. It defines a global mean efficiency for the manipulation, with the limit that any average may have. If the manipulation has a duration $T$, then:
\begin{equation}
\label{meanShrekNumber}
Sh = \frac1T \int_0^T iSh(t) dt
\end{equation}
\noindent The mean Shrek number associated to normal ventilation in a normal lung is about $0.2$.

It is important to highlight that Shrek number is only an indicator which aggregates many information, thus its use is correlated to an understanding of the physical phenomena involved in chest physiotherapy and to an understanding of its limitation due to the fact that it is an average.

\begin{figure}[h!]
\includegraphics[height=5cm]{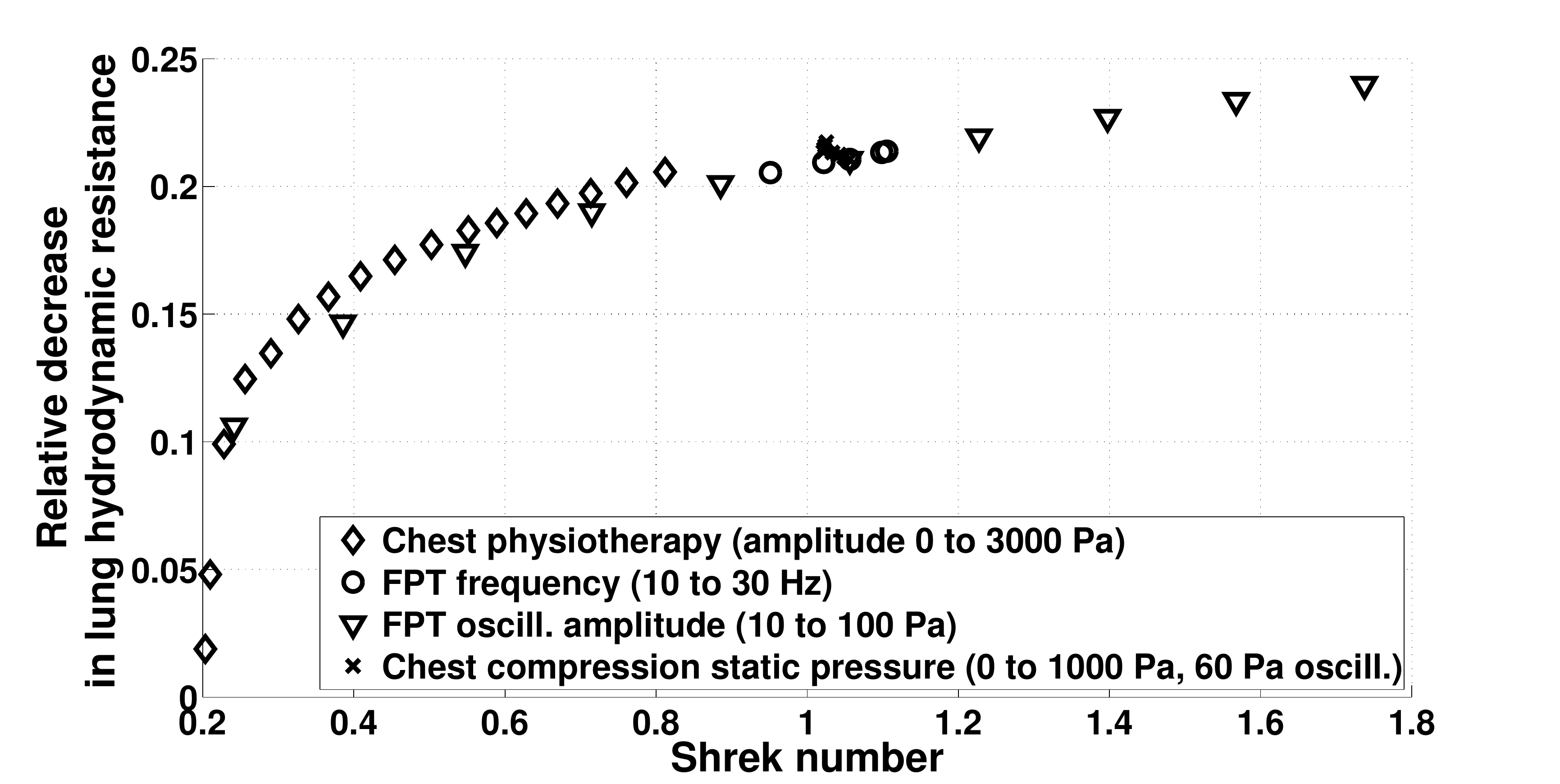}
\caption{Relative reduction of lung hydrodynamic resistance after $230$ seconds manipulations (model predictions) as a function of the time averaged Shrek number.}
\label{shrekExp}
\end{figure} 

On figure \ref{shrekExp}, we compare the efficiency in term of hydrodynamic resistance decrease for all the manipulations we studied in the previous sections. Most particularly, we can see that two manipulations with the same Shrek number have globally the same effect on hydrodynamic resistance. As expected, modulating the static pressure in chest compression techniques leads to very similar Shrek numbers, while manual and focused pulses techniques are able to cover a wide range of Shrek number by modulating their respective pressure amplitude.

\subsection{Comfort number: a measure for chest physiotherapy comfort}
\label{weasel}

The instantaneous comfort number of a manipulation is the lung tissue pressure shift relatively to normal ventilation. If at a given time of a manipulation, the volume of the lung is $V_L$ while it would have been $V_L^{ventil}$ without the manipulation, then the instantaneous comfort number is:
\begin{equation}
iCom = \left| \frac{P_{tissue}(V_L) - P_{tissue}(V_L^{ventil})}{P_{tissue}(V_L^{ventil})} \right|
\end{equation}
The instantaneous comfort number is zero if only ventilation occurs. The higher the number, the larger are the mechanical constraints in the lung tissues. Notice that this number only measures the mechanical stress inside the lung and is not directly related to its hydrodynamic resistance. For example if lung tissues are suffering high negative pressure because of an high inflation, the associated lung hydrodynamic resistance is probably low but mechanical stress is high. 

The comfort number of a manipulation with duration $T$ is the mean value computed over the whole manipulation,
\begin{equation}
Com = \frac1T \int_0^T iCom(t) dt
\end{equation}
As a consequence, a strong pressure applied for a short time can lead to similar comfort number as a moderate or even small pressure applied for a long time.

\begin{figure}[h!]
\includegraphics[height=7cm]{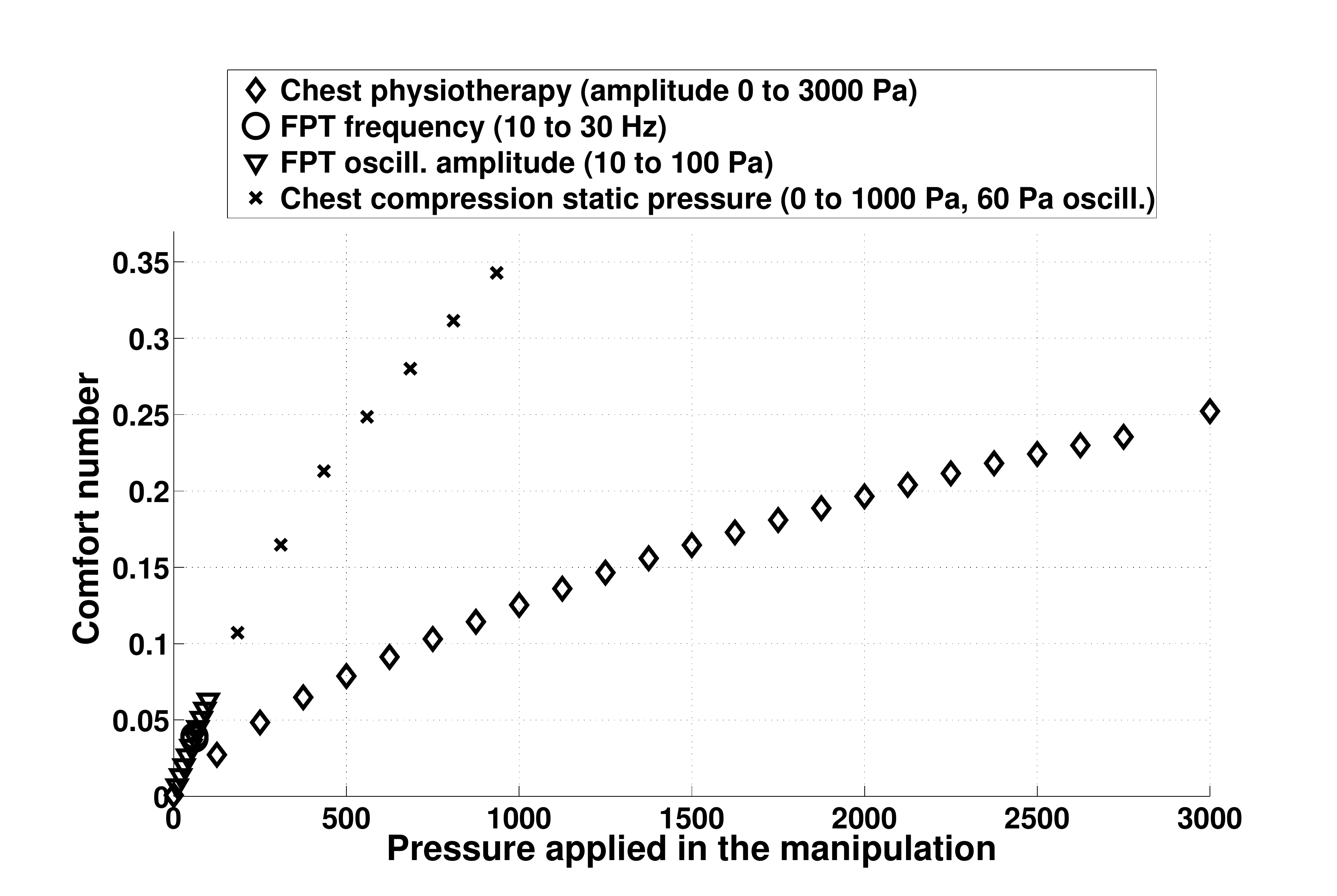}
\caption{Comfort number computed from model predictions for the different chest physiotherapy techniques mimicked in this study. The higher the comfort number, the larger are the stress in the lung tissues. The comfort number for normal ventilation is zero.}
\label{donkeyExp}
\end{figure} 

On figure \ref{donkeyExp}, the comfort number for different idealized chest physiotherapy techniques studied in this paper are plotted. Chest compression techniques with moderate static pressure are associated to high comfort numbers, mainly because the static pressure is applied constantly during the manipulation. Focused pulses technique are associated to low comfort numbers, because they are applying low pressures on chest. Notice that comfort number is only related to the net compressive forces induced into the lung by the manipulations, comfort number is not able to measure the role the frequency of pressure application may play on the patient comfort.

We recall that in our model the pressure is applied onto the whole chest. Thus in the case of manual chest physiotherapy,  comfort number has to be adapted to reflect the fact that the pressure is applied only on a subpart of the chest. In a first approximation, it can be done by weighting the pressure used to compute comfort number with the ratio between the surface on which the pressure is applied over the whole chest area.

\subsection{Model limitations}

Because the model described in this work contains the biophysical phenomena we identified as major in chest physiotherapy ("minimal" model), we are confident that it is able to predict main behaviors of chest physiotherapy. However, our predictions can only be qualitative or comparative since the biophysical phenomena included are simplified versions of the "real" phenomena and are based on static mechanics. For example, secretions are modeled by a simple Bingham fluid, or secretions thickness is assumed homogeneous in a bronchus. Moreover, quantitative predictions would need to include in the model some other phenomena which, although not strongly driving the system, might play a role on the quantitative data. Thus, any interpretation based on the results of our model has to be made in the frame of the model's hypotheses.

Our model is intrinsically non-linear. Non linearity in the model comes from two different aspects. First, in the way we deal with the mechanics of the lung and thorax. For this aspect, our model does not rely on a modeling process but instead on -non linear- data about lung mechanical inner pressure and about lung volumes (pressure curves from Agostoni et al. \cite{agostoni_static_2011}). Secondly, non linearity arises from the fluid mechanics of mucus which is based on Bingham fluid. Bingham fluid flows only if inner shear stress is above a threshold. Although non linear, secretion mechanics remains a basic simplification of real secretion mechanics, which behavior may be highly dependent on the local physical and biological environment. Air fluid mechanics in our model is linear, and we ignored inertia and turbulence which may break the homogeneity of flows distribution in the tree, even if the tree has symmetric branching \cite{mauroy_interplay_2003}. 

Asymmetric flow distribution can also be a consequence of an asymmetry in the geometry, either induced by asymmetric branching or by asymmetric secretions distribution in the tree. This last phenomenon may play a crucial role when a specific bronchus is more occluded by secretions than its neighbors: air flows might be redistributed towards the lowest resistive bronchial pathways and possibly miss the bronchi where secretions quantity is the highest. Chest physiotherapists are able to counteract this phenomena by applying pressures in a non homogeneous manner on the thorax. By assuming in our model that pressure is applied homogeneously on the thorax, none of these effects can be predicted with our current model.

Chest physiotherapy is highly patient dependent: manipulation are based on real-time response to the praticien feeling (hand feeling, noises, mouth air flows, etc.). The data we used in this model for secretions properties \cite{lai_micro-_2009}, pressure-volume relationships from Agostoni et al \cite{agostoni_static_2011} and lung geometry and mechanics \cite{lambert_computational_1982} reflect either mean or specific behaviors and cannot be applied as such to a specific patient. It has to be clear that the predictions of our model are general predictions that cannot be mapped as such to any patient. Again, careful interpretations in the frame of the model hypotheses have to be made before reaching any specific conclusion.

We proposed two numbers, the Shrek number that measures an efficiency for the manipulation and the comfort number that measures a degree of compressive discomfort correlated to a manipulation. Both these numbers have been tested with our idealized model. They are however not validated in a medical way and there significance remains as of today an hypothesis. In this paper we propose these numbers as candidates for evaluating chest physiotherapy manipulations, but clinical measurements have to be made before any medical applications of these numbers are made. This leads to another question: how to evaluate quantitatively the values for these numbers in a clinical frame. Concerning Shrek number, we proposed a formula that allows to compute its value with typical medical data, see equation (\ref{shrekNumber2}) and appendix \ref{approxShrek}. Concerning comfort number, it is necessary to get information about patient pressure-volume lung curves, which may be difficult in the frame of everyday consultations. This may be solved by assuming that patients pressure-volume lung curves have globally the same shapes and can be fitted through the measurement of a few simple physiological parameters at the beginning of a manipulation. This aspect is however out of the scope of this paper since it can be made only in a clinical frame.

\subsection{Conclusion}

We developed a model of the thorax and lung that is able to link the pressure on the thorax with the air and secretions motion in the airways. The goal of this model is to give indications on how chest physiotherapy may interact with the biomechanics of the thorax and of the lung. The model we built in this work is based on core phenomena that are involved in chest physiotherapy and can predict qualitative behaviors and compare different techniques, in the limitations of the model's hypotheses. 

Our results indicate that manipulations need to overcome secretions threshold in order to be able to mobilize secretions. Our model shows that manipulations main effect is to reduce the hydrodynamic resistance of the airway tree by secretions redistribution in the tree, eventually reaching a distribution that is not anymore affected by the manipulation. This effect means that patient breathing might be instantly improved by the manipulation. The second main effect is mucus expectoration, which is also another mean to improve the status of the patient. Mucus expectoration predicted by our model is non negligible only for pressures high enough, and also, in the case of HFCWO, for frequencies high enough.

All these phenomena are driven by the fact that the motor for secretions motion is the shear stress transmitted by air to the secretions. In a bronchus the shear stress is proportional to the air flow amplitude and to the inverse of the bronchus radius. Since air flow is created by applying pressure on the chest, the same method that is used to create the air flow will increase the hydrodynamic resistance of the airway tree, thus negatively affecting the air flow amplitude. Consequently, at some point, using a higher pressure will not really improve the efficiency of the manipulation, as shown on figure \ref{rp2}. This phenomenon is well known by chest physiotherapists and is also at the origin of forced expiration curve shape \cite{lambert_model_2004}.

Finally, we proposed two numbers to measure the efficiency of the manipulations: the Shrek number and the comfort number. They are both performing well in the case of our idealized manipulations. They need however to be validated with clinical studies. 

\section*{Acknowledgments}
The authors would like to thank RespInnovation SAS and the CNRS PEPS INSIS funding program for supporting this work.

\bibliographystyle{abbrv}
\bibliography{biblio}

\newpage
\appendix

\section{Computation of pressure drop per unit length - function $F$}
\label{AppF}

In this appendix, we describe how we compute the pressure drop per unit length in a branch, $C = \partial p / \partial z = F(\Phi_a, S_a, S_b)$, see equation (\ref{fm3}). The way we compute the function $F$ is identical to that used in \cite{mauroy_toward_2011}.  The expression of $F$ is a consequence of air fluid dynamics and depends on the mucus states.

All calculations take place in a branch of the tree whose radius is $r_b = \sqrt{S_b/\pi}$ and air lumen area radius is $r_a = \sqrt{S_a/\pi}$. The yield stress of mucus is $\sigma_0$ and its viscosity when it flows is $\mu_m$. Mucus is solid on the range of radius $[r_a, r_0]$ and liquid on the range $[r_0,r_b]$. If $r_0 \geq r_b$ then mucus is fully solid and if $r_0 \leq r_a$ then mucus is fully liquid. The yield radius $r_0$ is equal to $\left| \frac{2 \sigma_0}{C} \right|$, see \cite{mauroy_toward_2011}. Air viscosity is $\mu_a$.

We call $C=\partial p / \partial z$ the pressure drop per unit length in the branch. The air flow in the branch is known and is equal to $\Phi_a$. Finally, we call $s$ the sign of $C=\partial p / \partial z$ ($s=1$ if $C\leq 0$ and $s=-1$ if $C < 0$).

Because of developed flow and axi-symmetry hypotheses, fluid dynamics equations reduce to, see \cite{mauroy_toward_2011}:
$$
\text{for } r \in [0,r_a], \ -\frac{1}{r}\frac{\partial}{\partial r} \left( r \Sigma_{zr} \right) + \frac{\partial p}{\partial z} = 0 \ \text{ with } \ \Sigma_{zr} = \mu_a \frac{\partial v}{\partial r}
$$
By integration on $r$, the previous equation becomes: $\mu_a \frac{\partial v}{\partial r} = \frac{C r}{2}$.

The air flow in a branch is equal to: $\Phi_a = 2 \pi \int_0^{r_a} v(r) r dr$ and, integrating by parts, it becomes:
$$
\Phi_a = 2 \pi \left( v(r_a) \frac{r_a^2}{2} - \frac{C}{4 \mu_a} \frac{r_a^4}{4}\right)
$$
where $v(r_a)$ is the velocity of the air/mucus interface. This velocity depends on the state of mucus.

\begin{description}
\item[First state] mucus is liquid between the branch wall located at $r_b$ and the radius $r_0$ and solid elsewhere (case $r_a<r_0<r_b$). In that case $v(r_a)=v(r_0)$ since mucus is solid between $r_0$ and $r_a$. Then $v(r_a) = v(r_0) = -\frac{C}{4 \mu_m} (r_b-r_0)^2$. Replacing $v(r_a)$ with the previous expression in the expression of $\Phi_a$ and using the fact that $r_0=|2\sigma_0/C|$, then $C$ is the solution of a second degree polynomial. The two solutions are:
$$
C_1=\frac12 \frac{8 \pi r_a^2 \mu_a r_b \sigma_0-8 \phi_a \mu_a \mu_m+
4 \sqrt{-8 \pi r_a^2 \mu_a^2 r_b \sigma_0 \phi_a \mu_m+
4 \phi_a^2 \mu_a^2 \mu_m^2-2 \pi^2 r_a^6 \mu_a \sigma_0^2 \mu_m}}
{\pi r_a^4 \mu_m+2 \pi r_a^2 \mu_a r_b^2}
$$
and
$$
C_2=\frac12 \frac{8 \pi r_a^2 \mu_a r_b \sigma_0-8 \phi_a \mu_a \mu_m
-4 \sqrt{-8 \pi r_a^2 \mu_a^2 r_b \sigma_0 \phi_a \mu_m+4 \phi_a^2 
\mu_a^2 \mu_m^2-2 \pi^2 r_a^6 \mu_a \sigma_0^2 \mu_m}}
{\pi r_a^4 \mu_m+2 \pi r_a^2 \mu_a r_b^2}
$$
We have two possible pressure drops per unit length which gives two possible values for $r_0=|2\sigma_0/C|$, however they are easily discriminated since one only at a time can check $r_b<r_0<r_a$.\\
\item[Second state] mucus is liquid everywhere, i.e. $r_0<r_a$. Then, $v(r_a) = \frac{C}{4 \mu_m}(r_a-r_b)(r_a+r_b-2 r_0)$. Mixing with $\Phi_a$ expression leads to
$$
C=-\frac{8 \mu_a}{\pi r_a^2} 
\frac{\phi_a \mu_m-\pi r_a^3 \sigma_0+\pi r_a^2 r_b \sigma_0}
{r_a^2 \mu_m+2 r_a^2 \mu_a-2 \mu_a r_b^2}
$$

\item[Third state] mucus is completely solid ($r_0 > r_a$). In that case $v(r_a)=0$ and,
$$
C=-\frac{16 \mu_a \phi_a}{2 \pi r_a^4}
$$

\end{description}

Th function $F$ is built thanks to these formulas for $C$. To discriminate between the different possibilities and know the mucus state, we compute for each case the value(s) of $C$ and compute its associated radius $r_0=|2\sigma_0/C|$. The case is correct only if the position of the radius $r_0$ relatively to $r_a$ and $r_b$ is compatible with the case hypotheses.

\section{Computation of mucus flow - function $G$}
\label{AppG}

As for $F$, the computation of the mucus flow in a branch $\Phi_m = G(C,S_a,S_b)$ depends on the state of mucus, see equation (\ref{mm2}). The way we compute the function $G$ is identical to that used in \cite{mauroy_toward_2011}. All calculations take place in a branch of the tree whose radius is $r_b = \sqrt{S_b/\pi}$ and air lumen area radius is $r_a = \sqrt{S_a/\pi}$. The yield stress of mucus is $\sigma_0$ and its viscosity when it flows is $\mu_m$. Mucus is solid on the range of radius $[r_a, r_0]$ and liquid on the range $[r_0,r_b]$. If $r_0 \geq r_b$ then mucus is fully solid and if $r_0 \leq r_a$ then mucus is fully liquid. The yield radius $r_0$ is equal to $\left| \frac{2 \sigma_0}{C} \right|$, see \cite{mauroy_toward_2011}.

Fluid dynamics of mucus stands in the range $[r_a,r_b]$ and reduces to 
$$
\begin{array}{ll}
\frac{\partial v}{\partial r} = 0 & \ \text{where mucus is solid (typically in $[r_a,r_0]$)}\\
\Sigma_{zr} = \frac{C r}{2} = \sigma_0 + \mu_m \frac{\partial v}{\partial r} & \ \text{where mucus is liquid (typically in $[r_0,r_b]$)}\\
\end{array}
$$

The flow of mucus is given by
$$
\Phi_m = 2 \pi \int_{r_a}^{r_b} v(r) r dr 
$$

Solving fluid dynamics equations for air and mucus then leads to analytical expressions for mucus flow in the branch:
\begin{description}
\item[First case] mucus is liquid between the branch wall located at $r_b$ and the radius $r_0$ and solid elsewhere (case $r_a<r_0<r_b$). In that case $v(r_a)=v(r_0)$ since mucus is solid between $r_0$ and $r_a$. Then $v(r_a) = v(r_0) = -\frac{C}{4 \mu_m} (r_b-r_0)^2$, and
$$
\Phi_m =2 \pi \frac{v(r_a)}{2} (r_s^2-r_a^2)- 2 \pi \frac{C}{16 \mu_m} (r_b^2-r_s^2)^2
$$
\item[Second case] mucus is liquid everywhere, i.e. $r_0<r_a$. Then, $v(r_a) = \frac{C}{4 \mu_m}(r_a-r_b)(r_a+r_b-2 r_0)$, and 
$$
\Phi_m=- 2 \pi \frac{C}{16 \mu_m} (r_b^2-r_a^2)^2
$$
\item[Third case] mucus is completely solid ($r_0 > r_a$). In that case $v(r_a)=0$ and $\Phi_m = 0$.

\end{description}
The value of $r_0$ relatively to $r_a$ and $r_b$ allows to find in which case we are and then to compute the function $\Phi_m = G(C,S_a,S_b)$.

\section{Numerics}
\label{numerics}

The mathematical problem arising from the mechanics are solved using numerical computations. If $v=(v_i)_i$ and $w=(w_i)_i$ are two vectors then we define the product $v \star w$ as the vector $(v_i w_i)_i$. If $A$ is a matrix $n \times n$ and $v$ a vector of length $n$ then the matrix-vector product is written $A.x$.
 The equations are rewritten in a vectorial form using matrices-vector product $.$ and vector-vector product $ \star$:
$$
\left\{
\begin{array}{l}
S_b = H(P_{ext},P_{air})\\
\frac{d S_a}{dt} \star L_b = A.\Phi_a - n_{alv} \star \Phi_{alv}\\
C = F(\Phi_a, S_a, S_b)\\
P_{air} =  L.C\\
\left(\frac{d S_b}{dt} - \frac{d S_a}{dt}\right) \star L_b = |\Phi_{m,I}| - |\Phi_{m,O}|\\
\Phi_{m,O} = G(C, S_a, S_b)\\
\Phi_{m,I} = \frac12 \max\left(0,U.\Phi_{m,O}\right) - 2 \min\left(0,D.\Phi_{m,O}\right)
\end{array}
\right.
$$

Each vector $X$ has its coordinates denoted $X^{(z)}$, for example $S_b = (\Sb)_z$; the number of vector components is equal to the number of generations considered in the model ($23$). $A$, $L$, $U$ and $D$ are square matrices, see appendix \ref{matr} for details on their expressions. $G$ and $F$ are applied to vectors in this way: $G(C, S_a, S_b)=( G(\C, \Sa, \Sb)  )_z$ and $F(\Phi_a, S_a, S_b)=( F(\FaO, \Sa, \Sb)  )_z$.

The equations lead to a non linear vectorial ordinary differential system. It is solved in a full  implicit way. Given an initial condition at time $0$ and the time-dependence of the tissue pressure $t \rightarrow P_{tissue}(t)$, the system is discretized in this way:
$$
\left\{
\begin{array}{l}
S_b(t_n) = H(P_{ext}(t_n),P_{air}(t_n))\\
\Phi_a(t_n) = A^{-1} \left(\frac{S_a(t_n) - S_a(t_{n-1})}{t_n-t_{n-1}} \star L_b \right)\\
C(t_n) = F(\Phi_a(t_n), S_a(t_n), S_b(t_n))\\
P_{air}(t_n) =  L.C(t_n)\\
S_b(t_n) - S_a(t_n) = S_b(t_{n-1}) - S_a(t_{n-1}) +  (t_n - t_{n-1})  (|\Phi_{m,I}| - |\Phi_{m,O}|) \star L_b^{-1}\\
\Phi_{m,O}(t_n) = G(C(t_n), S_a(t_n), S_b(t_n))\\
\Phi_{m,I}(t_n)= \frac12 \max\left(0,U.\Phi_{m,O}(t_n)\right) - 2 \min\left(0,D.\Phi_{m,O}(t_n)\right)
\end{array}
\right.
$$

Concretely, we are able to eliminate all -vectorial- variables except $C(t_n)$ and $S_a(t_n)$, thus this system reduces to two non linear vectorial equations:
$$
\left\{
\begin{array}{l}
S_a(t_n) = \mathcal{F}_1 \left( S_a(t_n), C(t_n), S_a(t_{n-1}) , S_b(t_{n-1}) \right)\\
C(t_n) = \mathcal{F}_2 \left( S_a(t_n), C(t_n),  S_a(t_{n-1}) \right)
\end{array}
\right.
$$

The system is solved in this way at each time step $n$: 
\begin{itemize}
\item in a first step, we build a numerical function $\mathcal{C}_n(S)$ such that $\mathcal{C}_n(S) = \mathcal{F}_2 \left( S, \mathcal{C}_n(S),  S_a(t_{n-1}) \right)$. The function $\mathcal{C}_n$ is well defined since  $\mathcal{C}_n^{(z)}$ it is a strictly decreasing function of $S^{(z)}$. We use a Newton method to compute $\mathcal{C}_n$, initialized by $C_{init} = L.G(\Phi, S_a(t_n), \Sc (P_{tissue}(t_n)-P_{air}(t_{n-1})))$ with $\phi = A^{-1} \left(\frac{S - S_a(t_{n-1})}{t_n-t_{n-1}} \star L_b \right)$.
\item in a second step, we solve the equation $S = \mathcal{F}_1 \left( S, \mathcal{C}(S), S_a(t_{n-1}) , S_b(t_{n-1}) \right)$, again using a Newton method initialized with the air lumen area of the previous time step: $S_{init}=S_a(t_{n-1})$.
\end{itemize}

We use a parallel Newton method implemented in $C++$ and $OpenMP$. All equations and variables are normalized.

\section{Computation of matrices for the problem vectorial formulation}
\label{matr}

In this appendix, we give the expression of the matrices $A$, $L$, $U$ and $D$ used in the mathematical formulation of the model. In this appendix, the letter $i$ and $j$ refer to generation index and belong to the set $\{0,1,2, ...,N-1\}$.

The matrix $L$ relates the pressure drop per unit length to the mean pressure inside the bronchi, see equation (\ref{fm4}): 
$$
L_{i,j} =
\left\{
\begin{array}{ccc}
0 & \text{if} & j<i\\
l_i/2 & \text{if} & j=i\\
 l_i & \text{if} & j>i
 \end{array}
 \right.
 $$

The matrix $U$ is used to compute the mucus flow when it goes up in the tree (toward the upper bronchi):
$$
U_{i,j} =
\left\{
\begin{array}{ccc}
0 & \text{if} & j \neq i+1\\
1 & \text{if} & j=i+1
 \end{array}
 \right.
 $$

The matrix $D$ is used to compute the mucus flow when it goes up in the tree (toward the upper bronchi):
$$
D_{i,j} =
\left\{
\begin{array}{ccc}
0 & \text{if} & j \neq i-1\\
1 & \text{if} & j=i-1
 \end{array}
 \right.
 $$

The matrix $A$ relates the volume change of a bronchus with the air flows coming from the two daughter bronchi: $A = Id - 2 U$. 

\section{Approximation of the Shrek number}
\label{approxShrek}

In this appendix, we detail how the Shrek number can be approximated with equation (\ref{shrekNumber}), starting from its definition in equation (\ref{shrekNumberDef}). First, we need to make the hydrodynamic resistance of an airway in generation $i$, $R_i = 8 \mu_a l_i / (\pi r_i^4)$ appear in the sum:
$$
Sh = \frac1N \sum_{i=0}^{N-1} \frac{4 \mu_a \Phi_a^{(i)}}{\pi r_i^3 \sigma_0} = \frac{4}{N \sigma_0} \sum_{i=0}^{N-1} \frac{r_i}{8 l_i} \frac{8 \mu_a l_i}{\pi r_i^4} \Phi_a^{(i)}
$$
Then, using the fact that the tree is symmetric, we have $\Phi_a^{(i)} = \Phi_a^{(0)}/2^i$, thus:
$$
Sh = \frac{1}{2 N \sigma_0} \sum_{i=0}^{N-1} \frac{r_i}{l_i} \frac{R_i}{2^i} \Phi_a^{(0)}
$$
Because $\Phi_a^{(0)}$ is the flow in the trachea, then it corresponds to the mouth flow $\Phi_a$. The ratio $\frac{r_i}{l_i}$ is approximated by $\frac16$ since it is the mean value measured in the lung \cite{tawhai_ct-based_2004}:
$$
Sh = \frac{\Phi_a}{12 N \sigma_0} \sum_{i=0}^{N-1} \frac{R_i}{2^i} 
$$
Finally, the sum $\sum_{i=0}^{N-1} \frac{R_i}{2^i}$ is exactly the equivalent hydrodynamic resistance $R_{aw}$ of our model of the lung with symmetric branching \cite{mauroy_optimal_2004, mauroy_influence_2010}. Finally:
$$
Sh = \frac{R _{aw} \Phi_a}{12 N \sigma_0} 
$$

\end{document}